\documentclass[12pt,letterpaper]{article}
\usepackage{amsmath,amssymb,array,calc,rotating,epsfig,psfrag,amscd, cite}

\makeatletter
\renewcommand\section{\@startsection {section}{1}{\z@}%
                                   {-3.5ex \@plus -1ex \@minus -.2ex}%nn
                                   {2.3ex \@plus.2ex}%
                                   {\normalfont\large\bfseries}}
\renewcommand\subsection{\@startsection{subsection}{2}{\z@}%
                                     {-3.25ex\@plus -1ex \@minus -.2ex}%
                                     {1.5ex \@plus .2ex}%
                                     {\normalfont\bfseries}}
\makeatother

\let\non\nonumber

\let\l=\lambda

\let\s=\sigma

\let\S=\Sigma

\newcommand{\bea}{\begin{eqnarray}}
\newcommand{\eea}{\end{eqnarray}}
\newcommand{\be}{\begin{equation}}
\newcommand{\ee}{\end{equation}}

\newcommand{\p}{\partial}

\newcommand{\C}[1]{$(\ref{#1})$}

\def\IZ{\relax\ifmmode\mathchoice
{\hbox{\cmss Z\kern-.4em Z}}{\hbox{\cmss Z\kern-.4em Z}}
{\lower.9pt\hbox{\cmsss Z\kern-.4em Z}} {\lower1.2pt\hbox{\cmsss
Z\kern-.4em Z}}\else{\cmss Z\kern-.4em Z}\fi}
\def\IR{\relax{\rm I\kern-.18em R}}

\def\one{{\hbox{ 1\kern-.8mm l}}}

\newlength{\bredde}
\def\slash#1{\settowidth{\bredde}{$#1$}\ifmmode\,\raisebox{.15ex}{/}
\hspace*{-\bredde} #1\else$\,\raisebox{.15ex}{/}\hspace*{-\bredde}
#1$\fi}

\newsavebox{\zzzbar}
\sbox{\zzzbar}
  {\setlength{\unitlength}{0.9em}
  \begin{picture}(0.6,0.7)
  \thinlines
  \put(0,0){\line(1,0){0.6}}
  \put(0,0.75){\line(1,0){0.575}}
  \multiput(0,0)(0.0125,0.025){30}{\rule{0.3pt}{0.3pt}}
  \multiput(0.2,0)(0.0125,0.025){30}{\rule{0.3pt}{0.3pt}}
  \put(0,0.75){\line(0,-1){0.15}}
  \put(0.015,0.75){\line(0,-1){0.1}}
  \put(0.03,0.75){\line(0,-1){0.075}}
  \put(0.045,0.75){\line(0,-1){0.05}}
  \put(0.05,0.75){\line(0,-1){0.025}}
  \put(0.6,0){\line(0,1){0.15}}
  \put(0.585,0){\line(0,1){0.1}}
  \put(0.57,0){\line(0,1){0.075}}
  \put(0.555,0){\line(0,1){0.05}}
  \put(0.55,0){\line(0,1){0.025}}
  \end{picture}}

\newcommand{\ena}{\end{eqnarray}}
\newcommand{\beqa}{\begin{eqnarray}}
\newcommand{\eeqa}{\end{eqnarray}}

\def\l{\lambda}

\def\s{\sigma}

\def\S{\Sigma}

\begin{document}
\begin{titlepage}

\begin{center}

%\hfill \today
%\hfill         \phantom{xxx}         

%\hfill HRI

\vskip 2 cm
{\Large \bf Some finite terms from ladder diagrams in three and four loop maximal supergravity}\\
\vskip 1.25 cm { Anirban Basu\footnote{email address:
    anirbanbasu@hri.res.in} } \\
{\vskip 0.5cm Harish--Chandra Research Institute, Chhatnag Road, Jhusi,\\
Allahabad 211019, India\\}

\end{center}

\vskip 2 cm

\begin{abstract}
\baselineskip=18pt

We consider the finite part of the leading local interactions in the low energy expansion of the four graviton amplitude from the ladder skeleton diagrams in maximal supergravity on $T^2$, at three and four loops. At three loops, we express the $D^8\mathcal{R}^4$ and $D^{10} \mathcal{R}^4$ amplitudes as integrals over the moduli space of an underlying auxiliary geometry.  These amplitudes are evaluated exactly for special values of the the moduli of the auxiliary geometry, where the integrand simplifies.  We also perform a similar analysis for the $D^8\mathcal{R}^4$ amplitude at four loops that arise from the ladder skeleton diagrams for a special value of a parameter in the moduli space of the auxiliary geometry. While the dependence of the amplitudes on the volume of the $T^2$ is very simple, the dependence on the complex structure of the $T^2$ is quite intricate. In some of the cases, the amplitude consists of terms each of which factorizes into a product of two $SL(2,\mathbb{Z})$ invariant modular forms. While one of the factors  is a non--holomorphic Eisenstein series, the other factor splits into a sum of modular forms each of which satisfies a Poisson equation on moduli space with source terms that are bilinear in the Eisenstein series. This leads to several possible perturbative contributions unto genus 5  in type II string theory on $S^1$. Unlike the one and two loop supergravity analysis, these amplitudes also receive non--perturbative contributions from bound states of three D--(anti)instantons in the IIB theory.

\end{abstract}

\end{titlepage}

%\pagestyle{plain}
%\baselineskip=18pt
% Try a wider skip
%\baselineskip=19pt
%%%%%%%%%%%%%%%%%%%%%%%%%%%%%%%%%%%%%%%%%%%%%%%%%%%%%%%%%%%%%%%%%%%%%%%%%%%%%%

\section{Introduction}

The effective action of string theory in various backgrounds contains valuable information about the dynamics of the theory. While the various interactions determine the S--matrices of the theory, they also yield detailed information about the various perturbative and non--perturbative U--duality symmetries of the theory. While constructing the effective action in generic backgrounds is difficult, certain terms in the effective action can be determined in compactifications that preserve maximal supersymmetry\footnote{For theories with self--dual field strengths, what we mean are covariant equations of motion.}. These BPS interactions have been studied in considerable detail, though not much is known about the non--BPS interactions~\cite{Green:1997tv,Green:1997as,Kiritsis:1997em,Green:1998by,Obers:1998fb,Green:1999pu,Green:2005ba,Berkovits:2006vc,Basu:2007ru,Basu:2007ck,Green:2008bf,Basu:2008cf,Green:2010kv,Green:2010wi,Basu:2011he,Basu:2013goa,Basu:2013oka,D'Hoker:2013eea,Gomez:2013sla,D'Hoker:2014gfa,Bossard:2014lra,Basu:2014hsa,Basu:2014uba,Pioline:2015yea,Wang:2015jna,Bossard:2015uga}. The BPS interactions satisfy several non--renormalization theorems, which the non--BPS ones do not.   

Among the several methods that have been used to analyze these various terms in the effective action, a pivotal role has been played by maximal supergravity. This is because of the ability to calculate multi loop amplitudes in maximal supergravity~\cite{Green:1982sw,Green:1997as,Bern:1998ug,Green:1999pu,Green:2005ba,Green:2008bf,Bern:2007hh,Bern:2008pv,Bern:2009kd,Bern:2010ue,Basu:2014hsa,Basu:2014uba}. The prototypical example that we shall be concerned with is the four graviton amplitude in maximal supergravity. Considering the local interactions and performing a derivative expansion, this amplitude encodes the moduli dependence of the $D^{2k}\mathcal{R}^4$ interactions ($k$ is an integer) in toroidal compactifications of M theory and type II theory. Of course the ultraviolet divergences of supergravity have to be regularized consistent with the U--duality symmetries of string theory to yield finite answers. For toroidal compactifications of $N=1, d=11$ supergravity on $T^d$ for $d \leq 2$, one gets exact answers for these amplitudes. For $d \geq 3$, there are additional contributions in M theory from membrane and five brane instantons which are not captured by the supergravity analysis, and this gives the partial answer. This method has proven particularly useful to obtain the exact forms of some terms in the effective action upto three loops in supergravity. Several non--local terms in the effective action can also be obtained from these supergravity amplitudes, though they have not been as well studied.     

The one loop supergravity amplitude contains local interactions of the form $D^{2k} \mathcal{R}^4$ for $k \geq 0$, while the two, three and four loop amplitudes contain these interactions for $k\geq 2$, $k \geq 3$ and $k\geq 4$ respectively. The precise structure of these amplitudes beyond four loops is not well understood, while it is expected that they should contribute to amplitudes for $k\geq 4$ as well~\cite{Bjornsson:2010wm,Bjornsson:2010wu}. This is because the $\mathcal{R}^4, D^4 \mathcal{R}^4$ and $D^6\mathcal{R}^4$ interactions are $1/2,1/4$ and $1/8$ BPS respectively and satisfy non--renormalization theorems. Thus they receive perturbative contributions from only finite genera in string perturbation theory which are encoded in the moduli dependence of the supergravity loop amplitudes and hence they do not receive contributions beyond a few loops. On the other hand the $D^{2k} \mathcal{R}^4$ interactions for $k \geq 4$ are non---BPS and do not satisfy any non--renormalization theorems. Hence they are generically expected to receive perturbative contributions from all genera in string perturbation theory leading to the qualitative difference in the structure of the supergravity amplitudes beyond four loops.            

These supergravity amplitudes at one and two loops have an interesting underlying structure. This can be easily observed when they are obtained from the first quantized superparticle formalism~\cite{Berkovits:2001rb,Anguelova:2004pg,Dai:2006vj,Green:2008bf,Bjornsson:2010wm,Bjornsson:2010wu}. To see this we simply remove the external particles and focus on the underlying skeleton diagrams. The skeleton diagrams at one and two loops are given by figure 1 and figure 2 respectively.    

\begin{figure}[ht]
\begin{center}
\[
\mbox{\begin{picture}(100,60)(0,0)
\includegraphics[scale=.45]{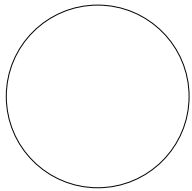}
\end{picture}}
\]
\caption{The one loop ladder skeleton}
\end{center}
\end{figure}

\begin{figure}[ht]
\begin{center}
\[
\mbox{\begin{picture}(100,60)(0,0)
\includegraphics[scale=.4]{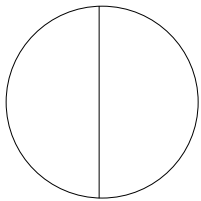}
\end{picture}}
\]
\caption{The two loop ladder skeleton}
\end{center}
\end{figure}

Thus at each loop order there is a unique skeleton diagram, and a given loop amplitude is obtained by attaching the external states represented as vertex operators to these skeleton diagrams. The integration over the links of these skeleton diagrams boils down to integration over the corresponding Schwinger parameters in the loop diagrams. While at one loop that is all that remains to be done, at two loops the three moduli of the ladder skeleton diagram can be very usefully expressed as the volume and complex structure of an auxiliary $T^2$~\cite{Green:1999pu,Green:2005ba}. Thus the measure and the lattice factor involving the Kaluza--Klein momenta are $SL(2,\mathbb{Z})$ invariant. However, it is noteworthy that the integrand is $SL(2,\mathbb{Z})$ invariant only for the $D^4\mathcal{R}^4$ interaction and not beyond. This is not entirely unexpected as the underlying auxiliary geometry does not follow from any symmetry principle. Nonetheless this geometrization of the two loop amplitude allows us to obtain compact expressions for the various couplings~\cite{Green:2008bf}. In particular, the ultraviolet divergences arise from the boundaries of moduli space and can be regularized. At one and two loop order the four graviton amplitude involves only massless $\varphi^3$ field theory propagators.   

The situation becomes more involved at three loops which has two distinct skeleton diagrams: the ladder and Mercedes skeletons given in figure 3. For the four graviton amplitude, the Mercedes skeleton contributes to the $D^{2k} \mathcal{R}^4$ interaction for $k \geq 3$, while the ladder skeleton contributes for $k \geq 4$. The integral over the six moduli of the Mercedes skeleton can be expressed as an integral over the moduli space of an auxiliary $T^3$ (volume and five shape moduli), with an $SL(3,\mathbb{Z})$ invariant measure and lattice factor~\cite{Basu:2014hsa}. While the integrand for the $D^6\mathcal{R}^4$ interaction is $SL(3,\mathbb{Z})$ invariant, this is no more the case for interactions at higher orders in the derivative expansion~\cite{Basu:2014uba}. Also the ultraviolet divergences arise from the boundary of moduli space and can be regularized. Thus there is a close analogy between the two loop ladder and three loop Mercedes skeleton diagram calculations. The three loop ladder skeleton contributions are more involved and will be discussed later. The four graviton amplitude involves only massless $\varphi^3$ field theory propagators for the ladder skeleton contributions, but not for the Mercedes skeleton contributions as the various integrands have non--trivial numerators.

\begin{figure}[ht]
\begin{center}
\[
\mbox{\begin{picture}(180,100)(0,0)
\includegraphics[scale=.4]{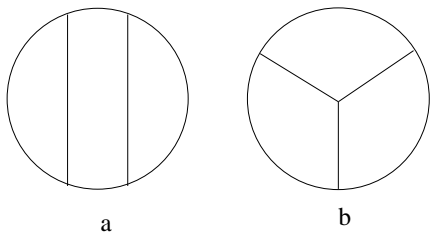}
\end{picture}}
\]
\caption{The three loop skeleton diagrams: (a) ladder and (b) Mercedes}
\end{center}
\end{figure}

The number of skeleton diagrams increases rapidly beyond three loops. For example, at four loops there are five such diagrams~\cite{Bern:2009kd} as depicted in figure 4. This gets more and more complicated at higher loops. 

\begin{figure}[ht]
\begin{center}
\[
\mbox{\begin{picture}(380,70)(0,0)
\includegraphics[scale=.55]{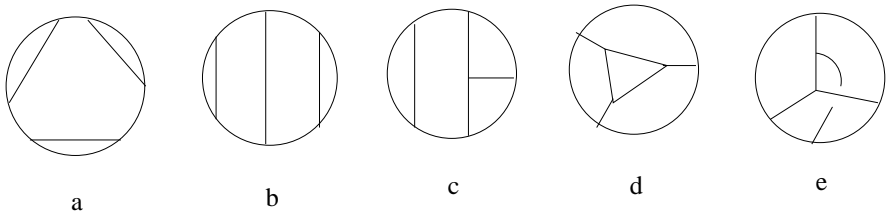}
\end{picture}}
\]
\caption{The four loop skeleton diagrams}
\end{center}
\end{figure}

Though there are new kinds of skeleton diagrams that arise at every order in the supergravity loop expansion, the ladder skeleton is universally present. These diagrams at three and four loops are the central content of the paper\footnote{We consider the $N=1, d=11$ theory compactified on $T^2$ so that the result is exact, but the analysis generalizes to arbitrary toroidal compactifications.}. Unlike the other skeleton diagrams at one, two and three loops, we do not have a detailed understanding of exactly what auxiliary geometry describes these diagrams (i.e. the solid whose group of large diffeomorphisms leaves the lattice factor and measure invariant). However, we do represent the three and four loop ladder skeleton integrals as integrals over the moduli space of some auxiliary geometry. In absence of the exact description of the geometry, we do not have a compact expression for the four graviton amplitude that follows from these geometries after integrating over moduli space. However, we do look at special values of the moduli of the auxiliary geometry where the integrand simplifies, and we can perform the analysis exactly. We justify the reason for this choice and argue why our analysis at these special values of the moduli gives us answers consistent with string theory. We shall see that the structure is intimately linked to supersymmetry. Thus we have a closed form expression for these amplitudes at certain special regions in the moduli space. These choices of values are not at the boundary of moduli space, though in evaluating the integrals for these values we have to regularize nested subdivergences that arise from the boundaries of moduli space. This yields finite answers which depend on $\mathcal{V}_2$, the volume of $T^2$ in the M theory metric, and $\Omega$, its complex structure. We also express them in terms of the type IIA and IIB theories compactified on $S^1$. Due to a lack of understanding of the details of the exact answer one would obtain on integrating over the full moduli space of the auxiliary geometry, we do not have a precise understanding of what our analysis exactly misses. However, we perform a similar analysis for the two loop supergravity amplitude where we can compare with the known exact answer. We show that our analysis yields the structure of the various ``source terms'' for the Poisson equations the various modular forms satisfy, which is a consequence of supersymmetry. We see that the results we obtain are consistent with known results in string perturbation theory. Hence we also expect our analysis to be useful in our understanding of the non--BPS interactions by providing us details about the structure of possible source terms which should exist for these amplitudes. 

In fact, for these values of the moduli the underlying geometries simplify, and hence are amenable to an exact analysis. For the case of the three and four loop amplitudes, they correspond to cases where the lattice factor and the measure are $SL(2,\mathbb{Z})$ invariant coming from auxiliary $T^2$s embedded in the auxiliary geometry.  The remaining moduli for these special values of moduli when they exist form a fibration over these $T^2$s in a very precise manner determined by the structure of the integrals. At three loop order, we perform our analysis for the $D^8\mathcal{R}^4$ and $D^{10} \mathcal{R}^4$ interactions which are the two leading local interactions in the derivative expansion. At four loop order, we perform a similar analysis for the $D^8\mathcal{R}^4$ interaction, again which is the leading local interaction. This analysis can be generalized to other local interactions at higher orders in the derivative expansion. Also the general structure of our analysis suggests an obvious generalization to all loop orders in supergravity, so far as the systematics of the ladder diagrams are concerned. This realizes an intricate interplay between various $SL(2,\mathbb{Z})$ subgroups of the lattice factors and the U--duality groups of the compactified theory.          

We first consider the three loop ladder diagram contributions to the $D^8\mathcal{R}^4$ and $D^{10} \mathcal{R}^4$ interactions. We then identify the moduli of the underlying geometry and perform the integrals for special values of the moduli. This involves an integral over the 5 Schwinger parameters of the ladder skeleton over a restricted region in moduli space. We next perform the same analysis for the four loop ladder diagram contribution to the $D^8\mathcal{R}^4$ interaction, which involves a restricted integral over 7 Schwinger parameters. Note that for these special values of the moduli, the amplitude contains both ultraviolet divergent as well as finite terms. The divergent terms have to be regularized to get finite answers. We consider only the finite terms in our analysis that do not need regularization.  Our analysis can also be generalized to obtain the divergent parts of the amplitude, which can be further regularized. 
Our analysis leads to several perturbative contributions at various genera in the type II theory compactified on $S^1$. For the $D^8\mathcal{R}^4$ interaction, this yields contributions unto genus 3, and upto genus 5 for the $D^{10}\mathcal{R}^4$ interaction. We examine how they match with known perturbative string results. Unlike the one and two loop supergravity analysis, these amplitudes also receive non--perturbative contributions from bound states of three D--(anti)instantons in the IIB theory which follow from the detailed non--perturbative structure of their $SL(2,\mathbb{Z})$ couplings.   

The exact expressions for the amplitudes we obtain for special values of the moduli of the auxiliary geometry have a simple dependence on $\mathcal{V}_2$ which follows from scaling. While for certain values of the moduli the amplitudes are independent of $\Omega$, the $\Omega$ dependence for other values is quite intricate. In fact, the amplitude comprises of terms which factorize into a product of two $SL(2,\mathbb{Z})$ invariant modular forms\footnote{The U--duality group in 9 dimensions is $SL(2,\mathbb{Z}) \times \mathbb{R}^+$.} in the Einstein frame. While one of these factors is given by a non--holomorphic Eisenstein series, the other factor splits into a sum of modular forms  each of which satisfies a Poisson equation on moduli space with source terms that are bilinear in the Eisenstein series. Solving these equations, we get various perturbative contributions to these amplitudes.

\section{The three loop ladder diagram contribution to the four graviton amplitude}

We consider $N=1, d=11$ supergravity compactified on $T^2$. The dimensionless volume of $T^2$ (in units of $4\pi^2 l_{11}^2$, where $l_{11}$ is the 11 dimensional Planck length) is $\mathcal{V}_2$ in the M theory metric, while its complex structure is $\Omega$.   Thus the metric of $T^2$ is given by
\be \label{defmet}
G_{IJ} = \frac{\mathcal{V}_2}{\Omega_2} \left( \begin{array}{cc} \vert \Omega \vert^2 & -\Omega_1 \\ -\Omega_1 & 1 \end{array} \right).\ee

The complete three loop four graviton amplitude is given by (we use the conventions of~\cite{Bern:2007hh,Bern:2008pv})
\bea \label{totcont}
\mathcal{A}_4^{(3)} = \frac{(4\pi^2)^3 \kappa_{11}^8}{(2\pi)^{33}}\sum_{S_3} \Big[ I^{(a)} + I^{(b)} + \frac{1}{2} I^{(c)} + \frac{1}{4} I^{(d)} + 2 I^{(e)} + 2 I^{(f)} + 4 I^{(g)} + \frac{1}{2} I^{(h)} + 2 I^{(i)}\Big] \mathcal{K} %&\equiv& \frac{(4\pi^2)^3 \kappa_{11}^8}{(2\pi)^{33}} I_3 \mathcal{K}. 
\non \\ \eea
where $S_3$ represents the 6 independent permutations of the external legs marked $\{1,2,3\}$ keeping the external leg $\{4\}$ fixed. The external momenta are directed inwards in all the loop diagrams and satisfy $k_i^2 =0, \sum_i k^M_i =0$, where $i=1,2,3,4$. Here $\mathcal{K}$ is the linearized approximation to $\mathcal{R}^4$ in momentum space, and $2\kappa_{11}^2 = (2\pi)^8 l_{11}^9$. $I^{(x)}$ refers to various loop diagrams, of which only diagrams $a,b$ and $d$ involve the ladder skeleton as given in figure 5, hence we only focus on them. The Mandelstam variables $S,T$ and $U$ are defined by $S = - G^{MN} (k_1+ k_2)_M (k_1 + k_2)_N, T = -G^{MN} (k_1+ k_4)_M(k_1 + k_4)_N$ and $U = - G^{MN} (k_1 + k_3)_M (k_1 + k_3)_N$.

\begin{figure}[ht]
\begin{center}
\[
\mbox{\begin{picture}(400,80)(0,0)
\includegraphics[scale=.55]{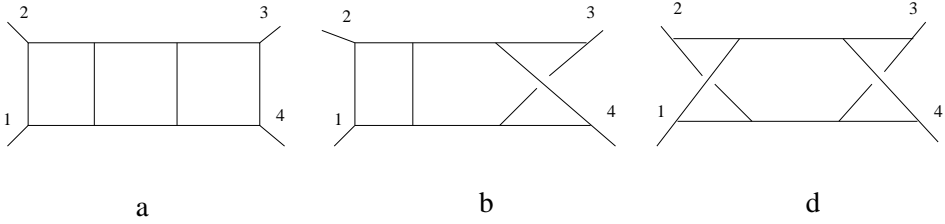}
\end{picture}}
\]
\caption{Three loop diagrams from the ladder skeleton}
\end{center}
\end{figure}

The numerators $N^{(x)}$ for the various integrands in the ladder loop diagrams (which have denominators given by massless $\varphi^3$ theory) are given by
\be N^{(a)} = N^{(b)} = N^{(d)} = S^4 .\ee

While we refer to each contribution mentioned in \C{totcont} as $I^{(x)}$, the total contribution after the sum over $S_3$ is referred to as $I^{(X)}$. We also denote
\be \S_n = S^n + T^n + T^n.\ee
We now consider the contribution coming from each diagram separately.

\subsection{The contribution from the planar diagram $a$}

In 11 uncompactified dimensions, the planar diagram $a$ contributes
\be \label{figa}
I^{(a)} = S^4\int  \frac{d^{11}p d^{11} q d^{11}r}{p^2 (p-k_2)^2 (p-k_1 -k_2)^2 q^2 (q+k_1 +k_2)^2 r^2 (r+k_3)^2 (r+k_3 + k_4)^2 (p+q)^2 (q+r)^2}.\ee
We now evaluate this integral, as well as all the other integrals, in the background $T^2 \times \mathbb{R}^{8,1}$. Thus the 11 dimensional loop momenta $p_M$, $q_M$ and $r_M$ decompose as $\{p_\mu,l_I/l_{11}\}$, $\{q_\mu, m_I/l_{11}\}$ and $\{r_\mu, n_I/l_{11}\}$ respectively where $p_\mu,q_\mu$ and $r_\mu$ are the 9 dimensional continuous momenta and the integers $l_I,m_I$ and $n_I$ ($I=1,2$) are the KK momenta along $T^2$. We introduce 10 Schwinger parameters $\s^i$ for the 10 propagators. Thus the product of the propagators in the compactified theory coming from the denominator of \C{figa} is given by  
\be \int_0^\infty \prod_{i=1}^{10} d\s^ie^{-\sum_{j=1}^{10}\s^j q_j^2} e^{-\Big((\s_1 +\s_2 + \s_3){\bf{l}}^2 +(\s_4 +\s_5){\bf{m^2}}  +(\s_6+ \s_7 +\s_8) {\bf{n^2}}  +\s_9 {\bf{(l+m)^2}} +\s_{10} {\bf{(m+n)^2}}\Big)/l_{11}^2}\ee
where
\be \label{defmom}q_j^2 = \{p^2,(p-k_2)^2,(p-k_1 - k_2)^2,q^2,(q+k_1 +k_2)^2,r^2,(r+k_3)^2,(r+k_3+k_4)^2,(p+q)^2,(q+r)^2\} \ee
and
\be {\bf{m}^2}\equiv G^{IJ} m_I m_J.\ee
Thus, compactifying on $T^2$ we have that
\bea \label{terma}I^{(a)} = \frac{S^4}{(4\pi^2 l_{11}^2 \mathcal{V}_2)^3} \int d^9 p \int d^9 q \int d^9 r \int_0^\infty \prod_{i=1}^{10} d\s^i e^{-\sum_{j=1}^{10}\s^j q_j^2} F_L (\s,\lambda, \rho,\s_9, \s_{10}),\eea  
where we have defined
\be \s = \s_1 +\s_2 +\s_3, \quad \lambda = \s_4 +\s_5, \quad \rho = \s_6 + \s_7 +\s_8.\ee
The lattice factor $F_L$ depends on only 5 independent Schwinger parameters and is given by
\bea \label{defF} F_L (\s,\lambda,\rho,\mu,\theta) = \sum_{l_I, m_I, n_I} e^{-G^{IJ}\Big( \s l_I l_J +\lambda m_I m_J +\rho n_I n_J +\mu (l+m)_I (l+m)_J +\theta (m+n)_I(m+n)_J  \Big)/l_{11}^2}\eea
which we simply denote $F_L$ for brevity\footnote{The specific order of the parameters is important, and is implicit in our discussions.}. Note that the Schwinger parameters $\s^i$ have dimensions $({\rm length})^2$.
We now define
\be w_1 = \frac{\s_1}{\s}, \quad w_2 = \frac{\s_1 +\s_2}{\s}, \quad u=\frac{\s_4}{\lambda}, \quad v_1 = \frac{\s_6}{\rho}, \quad v_2 = \frac{\s_6 +\s_7}{\rho}\ee
and hence
\be 0 \leq w_1 \leq w_2 \leq 1, \quad 0 \leq u \leq 1, \quad 0 \leq v_1 \leq v_2  \leq 1.\ee
This simplifies the expression for $I^{(a)}$ leading to
\bea I^{(a)} &=& \frac{S^4}{(4\pi^2 l_{11}^2 \mathcal{V}_2)^3} \int_0^\infty d\Upsilon \int_0^1 dw_2 \int_0^{w_2} dw_1 \int_0^1 du \int_0^1 dv_2 \int _0^{v_2} dv_1 \s^2 \lambda \rho^2 F_L \non \\ &&\times F^{(a)}_P (\s,\lambda,\rho,\mu,\theta,w_1,w_2,u,v_1,v_2),\eea 
where we have defined
\be d\s d\lambda d\rho d\mu d\theta= d\Upsilon. \ee
Thus the integrals over the 10 Schwinger parameters split into 5 ``radial'' integrals over $\s,\lambda,\rho,\mu,\theta$, and 5 ``angular'' integrals over $w_1,w_2,u,v_1,v_2$. The five radial integrals arise from the integration over the links of the ladder diagram and hence are universal for the three loop amplitude, while the angular integrals arise from integration over the insertion points of the vertex operators on the various links. Hence the structure of the angular integrals depends on specific diagrams.

Also $F^{(a)}_P (\s,\lambda,\rho,\mu,\theta,w_1,w_2,u,v_1,v_2)$ is given by
\be F^{(a)}_P (\s,\lambda,\rho,\mu,\theta,w_1,w_2,u,v_1,v_2) = \int d^9 p \int d^9 q \int d^9 r e^{-\sum_{j=1}^{10}\hat{\s}^j q_j^2},\ee 
where $q_j^2$ is given by \C{defmom} and 
\be \hat{\s}^j = \{\s_1, \s_2, \s_3, \s_4, \s_5, \s_6, \s_7, \s_8, \mu, \theta \}.\ee

We now evaluate the momentum integrals by using the relation
\bea &&\int d^9 p d^9 q d^9 r e^{-(A p^2 + B q^2 + C r^2 + D p \cdot q + E q \cdot r + p \cdot l_1 + q \cdot  l_2 + r \cdot l_3)} \non \\ &&= \frac{\pi^{27/2}}{\Delta^{9/2}} e^{[(BC - E^2/4)l_1^2 + AC l_2^2 + (AB-D^2/4) l_3^2 -CD l_1 \cdot l_2 - AE l_2 \cdot l_3 + DE l_1 \cdot l_3/2]/4\Delta}\eea
where $A,B,C >0$, and
\be \Delta = ABC - \frac{1}{4}(AE^2 + CD^2).\ee

We now consider the leading local contribution in the derivative expansion.
Thus, for the $D^8\mathcal{R}^4$ interaction, 
\bea I^{(a)}_{D^8\mathcal{R}^4} = \frac{\pi^{27/2} S^4}{4(4\pi^2 l_{11}^2 \mathcal{V}_2)^3} \int_0^\infty d\Upsilon   \frac{\s^2 \lambda \rho^2}{ \Delta_3^{9/2}} F_L  ,\eea 
where
\be \label{defdel}\Delta_3 (\s,\lambda, \rho,\mu,\theta) = \theta \lambda \s + \theta \lambda \mu +\theta \s \mu + \rho \s\lambda + \rho \s \mu+ \rho \theta \s + \rho \mu \lambda + \rho \mu \theta \ee
which we simply denote $\Delta_3$ for brevity\footnote{Again the specific order of the parameters is important, and is implicit in our discussions.}.
Thus
\bea \label{8-1}I^{(A)}_{D^8\mathcal{R}^4} = \frac{\pi^{27/2} \S_4}{2(4\pi^2 l_{11}^2 \mathcal{V}_2)^3} \int_0^\infty d\Upsilon \frac{\s^2 \lambda \rho^2}{\Delta_3^{9/2}}  F_L  .\eea 

Similarly for the $D^{10}\mathcal{R}^4$ interaction, we get that
\bea I^{(a)}_{D^{10}\mathcal{R}^4} = \frac{\pi^{27/2} S^4}{(4\pi^2 l_{11}^2 \mathcal{V}_2)^3} \int_0^\infty d\Upsilon \frac{\s^2 \lambda \rho^2}{\Delta_3^{9/2}}  F_L \Big[\frac{(\s+\rho+2\lambda)S}{48} + \frac{\mu\theta\s\rho}{72\Delta_3}(S+2T) \non \\ +\Big(\s\mu(\rho+\theta) +\rho\theta(\s+\mu)\Big)\frac{\lambda S}{48\Delta_3} \Big],\eea 
leading to
\bea \label{10-1}I^{(A)}_{D^{10}\mathcal{R}^4} = \frac{\pi^{27/2} \S_5}{24(4\pi^2 l_{11}^2 \mathcal{V}_2)^3} \int_0^\infty d\Upsilon \frac{\s^2 \lambda \rho^2}{\Delta_3^{9/2}}  F_L \Big[\s+\rho+3\lambda -\frac{\lambda^2 (\s+\mu)(\rho+\theta)}{\Delta_3} \Big].\eea 
The primary logic is the same for the other two diagrams, and so we shall omit several details.

\subsection{The contribution from the non--planar diagram $b$}

In 11 uncompactified dimensions, the non--planar diagram $b$ contributes
\be \label{figb}
I^{(b)} = S^4\int  \frac{d^{11}p d^{11} q d^{11}r}{p^2 (p-k_2)^2 (p-k_1 -k_2)^2 q^2 (q+k_1 +k_2)^2 r^2 (r+k_3)^2 (p+q)^2 (q+r)^2 (q+r-k_4)^2}.\ee
Compactifying on $T^2$ and proceeding as above, we get that
\bea \label{termb}I^{(b)} = \frac{S^4}{(4\pi^2 l_{11}^2 \mathcal{V}_2)^3} \int d^9 p \int d^9 q \int d^9 r \int_0^\infty \prod_{i=1}^{10} d\s^i e^{-\sum_{j=1}^{10}\s^j q_j^2} F_L (\s,\lambda, \rho,\s_8, \theta),\eea  
where we have defined
\be \s = \s_1 +\s_2 +\s_3, \quad \lambda = \s_4 +\s_5, \quad \rho = \s_6 + \s_7 , \quad \theta = \s_9 +\s_{10},\ee
and
\be  \label{defmomb}q_j^2 = \{p^2,(p-k_2)^2,(p-k_1 - k_2)^2,q^2,(q+k_1 +k_2)^2,r^2,(r+k_3)^2,(p+q)^2,(q+r)^2,(q+r-k_4)^2\} .\ee
We now define
\be w_1 = \frac{\s_1}{\s}, \quad w_2 = \frac{\s_1 +\s_2}{\s}, \quad u=\frac{\s_4}{\lambda}, \quad v = \frac{\s_6}{\rho}, \quad y = \frac{\s_9}{\theta}\ee
and hence
\be 0 \leq w_1 \leq w_2 \leq 1, \quad 0 \leq u,v,y \leq 1.\ee
Thus
\bea I^{(b)} &=& \frac{S^4}{(4\pi^2 l_{11}^2 \mathcal{V}_2)^3} \int_0^\infty d\Upsilon \int_0^1 dw_2 \int_0^{w_2} dw_1 \int_0^1 du \int_0^1 dv \int _0^1 dy \s^2 \lambda \rho \theta F_L  \non \\ &&\times F^{(b)}_P (\s,\lambda,\rho,\mu,\theta,w_1,w_2,u,v,y),\eea
where $F^{(b)}_P (\s,\lambda,\rho,\mu,\theta,w_1,w_2,u,v,y)$ is given by
\be F^{(b)}_P (\s,\lambda,\rho,\mu,\theta,w_1,w_2,u,v,y) = \int d^9 p \int d^9 q \int d^9 r e^{-\sum_{j=1}^{10}\hat{\s}^j q_j^2},\ee 
where $q_j^2$ is given by \C{defmomb} and 
\be \hat{\s}^j = \{\s_1, \s_2, \s_3, \s_4, \s_5, \s_6, \s_7,  \mu, \s_9, \s_{10} \}.\ee
Thus, for the $D^8\mathcal{R}^4$ interaction, 
\bea I^{(b)}_{D^8\mathcal{R}^4} = \frac{\pi^{27/2} S^4}{2(4\pi^2 l_{11}^2 \mathcal{V}_2)^3} \int_0^\infty d\Upsilon \frac{\s^2 \lambda \rho \theta}{\Delta_3^{9/2}}  F_L ,\eea
leading to
\bea \label{8-2}I^{(B)}_{D^8\mathcal{R}^4} = \frac{\pi^{27/2} \S_4}{(4\pi^2 l_{11}^2 \mathcal{V}_2)^3} \int_0^\infty d\Upsilon \frac{\s^2 \lambda \rho \theta}{\Delta_3^{9/2}}  F_L .\eea 

Similarly for the $D^{10}\mathcal{R}^4$ interaction,
\bea I^{(b)}_{D^{10}\mathcal{R}^4} &=& \frac{\pi^{27/2} S^4}{(4\pi^2 l_{11}^2 \mathcal{V}_2)^3} \int_0^\infty d\Upsilon \frac{\s^2 \lambda \rho \theta}{\Delta_3^{9/2}}  F_L \Big[ \Big(\s+\frac{3\lambda}{2}\Big)\frac{S}{6} -\frac{1}{4\Delta_3} \Big(\s^2 \{\theta(\lambda+\mu) +\rho(\lambda+\mu+\theta)\} \frac{S}{2} \non \\ &&+\frac{2}{3} \lambda (\s+\mu)(\rho+\theta)(\lambda+\frac{3\theta}{4})S -\rho\theta \{\mu(\lambda+\theta) +\s(\lambda+\mu+\theta)\}\frac{S}{2} \non \\ &&+\mu\s(\rho+\theta)(3\lambda+\theta)\frac{S}{3} + \theta(\s+\mu)(\lambda\rho-\lambda\theta +\theta\rho)\frac{S}{2} +\mu\theta\s(2\rho-\theta)\frac{S}{3}\Big)\Big],\eea
leading to
\bea \label{10-2}I^{(B)}_{D^{10}\mathcal{R}^4} = \frac{\pi^{27/2} \S_5}{12(4\pi^2 l_{11}^2 \mathcal{V}_2)^3} \int_0^\infty d\Upsilon \frac{\s^2 \lambda \rho \theta}{\Delta_3^{9/2}}  F_L \Big[\s+3\lambda -\frac{\lambda^2 (\s+\mu)(\rho+\theta)}{\Delta_3} \Big].\eea 

\subsection{The contribution from the non--planar diagram $d$}

In 11 uncompactified dimensions, the non--planar diagram $d$ contributes
\bea \label{figd}
I^{(d)} &=& S^4\int  \frac{d^{11}p d^{11} q d^{11}r}{(p-k_2)^2 (p-k_1 -k_2)^2 q^2 (q+k_1 +k_2)^2 r^2 (r+k_3)^2 (p+q)^2 (p+q-k_2)^2 } \non \\ &&\times \frac{1}{(q+r)^2 (q+r-k_4)^2}.\eea
Again compactifying on $T^2$ and proceeding as above, we get that
\bea \label{termd}I^{(d)} = \frac{S^4}{(4\pi^2 l_{11}^2 \mathcal{V}_2)^3} \int d^9 p \int d^9 q \int d^9 r \int_0^\infty \prod_{i=1}^{10} d\s^i e^{-\sum_{j=1}^{10}\s^j q_j^2} F_L (\s,\lambda, \rho,\mu, \theta),\eea  
where we have defined
\be \s = \s_1 +\s_2 , \quad \lambda = \s_3 +\s_4, \quad \rho = \s_5 + \s_6 , \quad \mu = \s_7 +\s_8, \quad \theta = \s_9 +\s_{10},\ee
and
\be  \label{defmomd}q_j^2 = \{(p-k_2)^2,(p-k_1 - k_2)^2,q^2,(q+k_1 +k_2)^2,r^2,(r+k_3)^2,(p+q)^2,(p+q-k_2)^2,(q+r)^2,(q+r-k_4)^2\} . \ee
We now define
\be w = \frac{\s_1}{\s}, \quad u = \frac{\s_3}{\lambda}, \quad v=\frac{\s_5}{\rho}, \quad y = \frac{\s_7}{\mu}, \quad z = \frac{\s_9}{\theta}\ee
and hence
\be  0 \leq w,u,v,y,z \leq 1.\ee
Thus we have that
\bea I^{(d)} &=& \frac{S^4}{(4\pi^2 l_{11}^2 \mathcal{V}_2)^3} \int_0^\infty d\Upsilon \int_0^1 dw \int_0^1 du \int_0^1 dv \int_0^1 dy \int _0^1 dz \s \lambda \rho \mu \theta F_L  \non \\ &&\times F^{(d)}_P (\s,\lambda,\rho,\mu,\theta,w,u,v,y,z),\eea
where $F^{(d)}_P (\s,\lambda,\rho,\mu,\theta,w,u,v,y,z)$ is given by
\be F^{(d)}_P (\s,\lambda,\rho,\mu,\theta,w,u,v,y,z) = \int d^9 p \int d^9 q \int d^9 r e^{-\sum_{j=1}^{10}\hat{\s}^j q_j^2},\ee 
where $q_j^2$ is given by \C{defmomd} and 
\be \hat{\s}^j = \{\s_1, \s_2, \s_3, \s_4, \s_5, \s_6, \s_7,  \s_8, \s_9, \s_{10} \}.\ee
Thus, for the $D^8\mathcal{R}^4$ interaction, 
\bea I^{(d)}_{D^8\mathcal{R}^4} = \frac{\pi^{27/2} S^4}{(4\pi^2 l_{11}^2 \mathcal{V}_2)^3} \int_0^\infty d\Upsilon \frac{\s \lambda \rho \mu \theta}{\Delta_3^{9/2}}  F_L ,\eea
leading to
\bea \label{8-3}I^{(D)}_{D^8\mathcal{R}^4} = \frac{2\pi^{27/2} \S_4}{(4\pi^2 l_{11}^2 \mathcal{V}_2)^3} \int_0^\infty d\Upsilon \frac{\s \lambda \rho \mu \theta}{\Delta_3^{9/2}}  F_L .\eea

Similarly for the $D^{10}\mathcal{R}^4$ interaction,
\bea I^{(d)}_{D^{10}\mathcal{R}^4} &=& \frac{\pi^{27/2} S^5}{(4\pi^2 l_{11}^2 \mathcal{V}_2)^3} \int_0^\infty d\Upsilon \frac{\s \lambda \rho \mu \theta}{\Delta_3^{9/2}}  F_L \Big[\frac{\s+\l}{2} -\frac{1}{4\Delta_3}\Big( \s(2\s+\mu) \{ \theta(\lambda+\mu) +\rho(\lambda+\mu+\theta)\} \non \\ &&+\lambda(\s+\mu)(\rho+\theta)(\frac{4\lambda}{3} -\mu+\theta) -\rho\theta \{\mu(\lambda+\theta) + \s(\lambda+\mu+\theta)\} \non \\ &&+\mu(\rho+\theta)\{\s(3\lambda-\mu +\theta)+\lambda\mu\} +\mu\theta\s(\rho-\theta) +\theta(\s+\mu)(\lambda(\rho-\theta) +\rho\theta)\Big) \Big].\eea
Though the integrand looks rather complicated, it can be simplified considerably using the symmetries of the ladder skeleton diagram discussed in appendix A. Note that the measure $d\Upsilon$, $\s\l\rho\mu\theta$, $F_L$ and $\Delta_3$ are all invariant under the symmetries of the ladder skeleton diagram. In fact several terms in the integral vanish using the transformations $P_i$ in \C{listtrans}.  
Thus in the integral, under $P_2$
\be \lambda(\s+\mu)(\rho+\theta)(\mu-\theta) \rightarrow -\lambda(\s+\mu)(\rho+\theta)(\mu-\theta) =0 .\ee
Also acting with $P_2$ on either of the terms, we see that
\be \mu^2 \s (\rho+\theta) - \theta^2 \rho(\s+\mu) =0.\ee
Similarly
\bea &&\lambda \mu^2 (\rho+\theta) - \lambda \theta^2 (\s+\mu) =0,\non \\ &&
\s\mu\{\theta(\lambda+\mu)+\rho(\lambda+\mu+\theta)\} -\rho\theta \{\mu(\lambda+\theta)+\s(\lambda+\mu+\theta)\}=0\eea
and
\be -3\mu \s\lambda (\rho+\theta) - \rho\theta\lambda (\s+\mu) \rightarrow  -4\mu \s\lambda (\rho+\theta)\ee
in the integral. 
This leads to
\bea I^{(d)}_{D^{10}\mathcal{R}^4} &=& \frac{\pi^{27/2} S^5}{(4\pi^2 l_{11}^2 \mathcal{V}_2)^3} \int_0^\infty d\Upsilon \frac{\s \lambda \rho \mu \theta}{\Delta_3^{9/2}}  F_L \Big[\frac{\lambda^2}{6\Delta_3} (\s+\mu)(\rho+\theta) +\frac{\lambda\theta\rho}{2\Delta_3} (\s+\mu) \Big],\eea
and hence
\bea \label{10-3}I^{(D)}_{D^{10}\mathcal{R}^4} &=& \frac{\pi^{27/2} \Sigma_5}{(4\pi^2 l_{11}^2 \mathcal{V}_2)^3} \int_0^\infty d\Upsilon \frac{\s \lambda \rho \mu \theta}{\Delta_3^{11/2}}  F_L \Big[\frac{\lambda^2}{3} (\s+\mu)(\rho+\theta) +\lambda\theta\rho (\s+\mu) \Big] \non \\ &=& \frac{\pi^{27/2} \Sigma_5}{2(4\pi^2 l_{11}^2 \mathcal{V}_2)^3} \int_0^\infty d\Upsilon \frac{\s \lambda \rho \mu \theta}{\Delta_3^{9/2}}  F_L \Big[\lambda -\frac{\lambda^2(\s+\mu)(\rho+\theta)}{3\Delta_3}  \Big]\eea
on using the symmetries of the ladder skeleton. 

Thus we see that for the various contributions, although the details are involved, the final expressions have reasonably simple forms.

\section{The $D^8\mathcal{R}^4$ and $D^{10}\mathcal{R}^4$ interactions from three loop ladder diagrams}

We now consider the total contribution coming from three loop ladder diagrams to the $D^8\mathcal{R}^4$ and $D^{10}\mathcal{R}^4$ interactions. Adding the various ladder diagram contributions in \C{8-1}, \C{8-2} and \C{8-3}, at $O(D^8\mathcal{R}^4)$ we get that
\bea \label{int1}I^{D^8 \mathcal{R}^4} \equiv I^{(A)}_{D^8\mathcal{R}^4} + I^{(B)}_{D^8\mathcal{R}^4} +\frac{1}{4} I^{(D)}_{D^8\mathcal{R}^4} =\frac{\pi^{27/2} \Sigma_4}{2(4\pi^2 l_{11}^2 \mathcal{V}_2)^3} \int_0^\infty \frac{d\Upsilon}{\Delta_3^{9/2}} F_L\Big(  \s^2 \lambda \rho^2 + 2 \s^2 \lambda \rho \theta +\s \lambda \rho \mu \theta \Big)  .\non \\ \eea
Similarly adding the various contributions in \C{10-1}, \C{10-2} and \C{10-3}, at 
 $O(D^{10}\mathcal{R}^4)$ we get that
\bea \label{int2}&&I^{D^{10} \mathcal{R}^4} \equiv I^{(A)}_{D^{10}\mathcal{R}^4} + I^{(B)}_{D^{10}\mathcal{R}^4} +\frac{1}{4} I^{(D)}_{D^{10}\mathcal{R}^4} \non \\  &&=\frac{\pi^{27/2} \Sigma_5}{24(4\pi^2 l_{11}^2 \mathcal{V}_2)^3} \int_0^\infty \frac{d\Upsilon}{\Delta_3^{9/2}}  F_L \Big[-\frac{\lambda^3\s\rho(\s+\mu)(\rho+\theta)}{\Delta_3} (\s\rho+2\s\theta+\mu\theta)   \non \\&& + \s\lambda\rho\{\s\rho(\s+\rho+3\lambda)+2\s\theta(\s+3\lambda)+3\lambda\mu\theta\}    \Big].\eea

\subsection{Expressing the $D^8\mathcal{R}^4$ and $D^{10}\mathcal{R}^4$ loop integrals in a $K_4$ invariant way}

It is very useful for our purposes to express the integrals in \C{int1} and \C{int2} in a manifestly $K_4$ invariant way. This is because we shall later describe the integrals as integrals over the moduli space of some auxiliary geometry. We want to do this reparametrization in a $K_4$ invariant way, which corresponds to the symmetry of relabelling the various loop momenta (compact and non--compact). To do so, we use the various symmetries of the ladder skeleton or equivalently the diamond graph as given in figure 10. How these symmetry transformations act on the 5 Schwinger parameters is given in \C{listtrans}. Now $F_L, \Delta_3$ and $d\Upsilon$ are $K_4$ invariant, hence we only need to write the remaining parts of the integrands in a $K_4$ invariant way.    

Hence we make the following replacements inside the integrals:
\bea &&\s\rho \rightarrow \frac{1}{2}(\s\rho+\mu\theta), \non \\ 
&& \s^2 \rho^2 \rightarrow \frac{1}{2}(\s^2\rho^2+\mu^2\theta^2), \non \\ 
&& \s\rho+2\s\theta+\mu\theta \rightarrow (\s+\mu)(\rho+\theta), \non \\ 
&& \s^2\rho\theta \rightarrow \frac{1}{4}\Big[(\s^2 +\mu^2)\rho\theta+(\rho^2+\theta^2) \s\mu\Big].
\non \\
 &&\s\rho(\s+\rho+3\lambda) +2\s\theta(\s+3\lambda)+3\lambda\mu\theta \rightarrow 3\lambda(\s+\mu)(\rho+\theta)\non \\ &&+\frac{1}{2} \Big[(\s^2 +\mu^2)(\rho+\theta)+(\rho^2 +\theta^2) (\s+\mu)\Big] \non \\&& = 4\lambda(\s+\mu)(\rho+\theta) -\Delta_3 +\frac{1}{2} (\s+\mu)(\rho+\theta)(\s+\mu+\rho+\theta).
\eea

Thus we get that
\bea \label{impeqn}
I^{D^8 \mathcal{R}^4} = \frac{\pi^{27/2} \Sigma_4}{4(4\pi^2 l_{11}^2 \mathcal{V}_2)^3} \int_0^\infty \frac{d\Upsilon}{\Delta_3^{9/2}} F_L \lambda \Big[  \Big(\s  \rho+\mu \theta\Big)^2 + \Big(\rho \theta (\s^2 +\mu^2) +\s\mu(\rho^2 +\theta^2)\Big)\Big] \non \\ = \frac{\pi^{27/2} \Sigma_4}{4(4\pi^2 l_{11}^2 \mathcal{V}_2)^3} \int_0^\infty \frac{d\Upsilon}{\Delta_3^{9/2}} F_L \lambda \Big[  \Big(\s  \rho-\mu \theta\Big)^2 + \Big(\rho \theta (\s +\mu)^2 +\s\mu(\rho +\theta)^2\Big)\Big]. \eea
and
\bea \label{impeqn2}
 I^{D^{10} \mathcal{R}^4} &=& \frac{\pi^{27/2} \Sigma_5}{48(4\pi^2 l_{11}^2 \mathcal{V}_2)^3} \int_0^\infty \frac{d\Upsilon}{\Delta_3^{9/2}} F_L \lambda (\s\rho+\mu\theta)\Big[  -\frac{\lambda^2}{\Delta_3} (\s+\mu)^2 (\rho+\theta)^2 \non \\ && +4\lambda(\s+\mu)(\rho+\theta) -\Delta_3 +\frac{1}{2}(\s+\mu)(\rho+\theta)(\s+\mu+\rho+\theta)\Big]. \eea

It is very convenient to perform a Poisson resummation to go from KK momentum basis to winding number basis, to separate the various ultraviolet divergences from the finite contributions. To do so, consider the lattice factor $F_L$. We write \C{defF} in a compact way as
\be \label{newdefF} F_L(\s,\lambda,\rho,\mu,\theta) = \sum_{k_{\alpha I}} e^{-G^{IJ}G^{\alpha\beta} k_{\alpha I}k_{\beta J}/l_{11}^2},\ee
where the KK integers $k_{\alpha I}$ are defined by $k_{\alpha I} = \{l_I,m_I,n_I\}$ for $\alpha =1,2,3$. In \C{newdefF}, the symmetric matrix $G^{\alpha\beta}$ ($\alpha,\beta =1,2,3$) has entries (of dimension $l_{11}^2$)
\be \label{inverse}G^{\alpha\beta} = \begin{pmatrix}
\s +\mu &\mu &0  \\
\mu & \lambda+\mu+\theta & \theta \\
0 & \theta & \rho+\theta
\end{pmatrix}.\ee    
Note that ${\rm det} G^{\alpha\beta} = \Delta_3(\s,\lambda,\rho,\mu,\theta)$. After Poisson resummation, we get that
\be \label{defF2} F_L (\s,\lambda,\rho,\mu,\theta)= \frac{(\pi l_{11}^2 \mathcal{V}_2)^3}{\Delta_3}\sum_{\hat{k}^{\alpha I}} e^{-\pi^2 l_{11}^2 G_{IJ}G_{\alpha\beta} \hat{k}^{\alpha I} \hat{k}^{\beta J}}, \ee
where the winding mode integers $\hat{k}^{\alpha I}$ are defined by $\hat{k}^{\alpha I} = \{\hat{l}^I,\hat{m}^I,\hat{n}^I\}$ for $\alpha =1,2,3$. Also the matrix $G_{\alpha\beta}$ is the inverse of the matrix $G^{\alpha\beta}$, and has entries of dimension $l_{11}^{-2}$.

It is illustrative to compare this structure with that of the Mercedes skeleton diagram. For that case, we had 6 Schwinger parameters which characterized the dual tetrahedron. By simply removing one of the 6 edges of the dual tetrahedron~\cite{Basu:2014hsa}, we get the diamond graph with 5 edges\footnote{In the notation of~\cite{Basu:2014hsa} this amounts to setting $\theta=0$ (see equation (6.129)).}. Thus we can think of the above dual geometry as a limiting case of the tetrahedral geometry.

We also want to redefine the Schwinger parameters after performing the Poisson resummation. So we define
\be \label{newS1}\hat\s = \frac{\s}{\Delta_3^{2/3}}, \quad \hat\lambda  = \frac{\lambda}{\Delta_3^{2/3}}, \quad \hat\rho  = \frac{\rho}{\Delta_3^{2/3}}, \quad \hat\mu  = \frac{\mu}{\Delta_3^{2/3}}, \quad \hat\theta  = \frac{\theta}{\Delta_3^{2/3}},  \ee
and so
\bea \label{newS2}\hat\Delta_3 (\hat\s,\hat\lambda,\hat\rho,\hat\mu,\hat\theta) =\hat\theta \hat\lambda \hat\s + \hat\theta \hat\lambda \hat\mu +\hat\theta \hat\s \hat\mu + \hat\rho \hat\s \hat\lambda + \hat\rho \hat\s \hat\mu+ \hat\rho \hat\theta \hat\s + \hat\rho \hat\mu \hat\lambda + \hat\rho \hat\mu \hat\theta  = \Delta_3^{-1} (\s,\lambda,\rho,\mu,\theta)\non \\ \eea
which we define $\hat\Delta_3$ for brevity.
Thus the variables $\hat\s,\hat\lambda,\hat\rho,\hat\mu$ and $\hat\theta$ have dimensions of $l_{11}^{-2}$.  The measure transforms in a very simple way as
\be d\Upsilon = \hat{\Delta}_3^{-10/3} d\hat\s d\hat\lambda d\hat\rho d\hat\mu d\hat\theta \equiv \hat{\Delta}_3^{-10/3} d\hat\Upsilon,\ee 
on using the relation
\be \Big( \s\frac{\p}{\p\s} + \lambda \frac{\p}{\p\lambda} +\rho \frac{\p}{\p\rho} +\mu \frac{\p}{\p\mu} +\theta \frac{\p}{\p\theta}\Big) \Delta_3 = 3 \Delta_3.\ee
Hence after Poisson resummation from \C{impeqn} we get that
\bea \label{impexp3}
I^{D^8 \mathcal{R}^4} = \frac{\pi^{21/2} \Sigma_4}{256} \int_0^\infty \frac{d\hat\Upsilon}{\hat\Delta_3^{7/6}}  \hat\lambda \Big[  \Big(\hat\s  \hat\rho-\hat\mu \hat\theta\Big)^2 +  \Big(\hat\rho \hat\theta (\hat\s +\hat\mu)^2 +\hat\s \hat\mu(\hat\rho +\hat\theta)^2\Big)\Big] \hat{F}_L, \eea
where we have defined the Poisson resummed lattice sum
\be \hat{F}_L \equiv \hat{F}_L (\hat\s,\hat\l,\hat\rho,\hat\mu,\hat\theta) = \sum_{\hat{k}^{\alpha I}} e^{-\pi^2 l_{11}^2 G_{IJ}G_{\alpha\beta} \hat{k}^{\alpha I} \hat{k}^{\beta J}}.\ee
Similarly from \C{impeqn2} we get that
\bea \label{impexp4} && I^{D^{10} \mathcal{R}^4} = \frac{\pi^{21/2} \Sigma_5}{3072} \int_0^\infty \frac{d\hat\Upsilon}{\hat\Delta_3^{11/6}} \hat{F}_L \hat\lambda (\hat\s\hat\rho+\hat\mu\hat\theta)\Big[  -\frac{\hat\lambda^2}{\hat\Delta_3} (\hat\s+\hat\mu)^2 (\hat\rho+\hat\theta)^2 \non \\ && +4\hat\lambda(\hat\s+\hat\mu)(\hat\rho+\hat\theta) -\hat\Delta_3 +\frac{1}{2}(\hat\s+\hat\mu)(\hat\rho+\hat\theta)(\hat\s+\hat\mu+\hat\rho+\hat\theta)\Big].\eea

These yield the unrenormalized $D^8\mathcal{R}^4$ and $D^{10} \mathcal{R}^4$ three loop amplitudes from ladder diagrams.  

\section{Mapping the Schwinger parameters to an auxiliary geometry}

We now express the $K_4$ invariant integrals in \C{impexp3} and \C{impexp4} which are integrals over the 5 Schwinger parameters in terms of integrals over the moduli space of an auxiliary geometry. Though we consider only these two interactions for simplicity, our analysis can be generalized for all the local interactions at arbitrary orders in the derivative expansion. 

We proceed by beginning with the results for the Mercedes diagrams. In the conventions of~\cite{Basu:2014hsa} the diamond graph is obtained by setting $\theta=0$, and hence $A_2=0$ (see equation (6.138)). Thus the surviving auxiliary geometry is a subspace of $T^3$. Needless to say, the $SL(3,\mathbb{Z})$ symmetry of $T^3$ which was very helpful in the analysis of the Mercedes diagrams, no more survives. This makes the analysis more complicated.

Thus we introduce 5 real dimensionless parameters $L,T_1,T_2,A$ and $V_3$ (thus $A_1=A$ in the conventions of~\cite{Basu:2014hsa}). 
We set the entries of $G^{\alpha\beta}$ to have dimension $l_{11}^2$ to match the entries in \C{inverse}. Thus
\be \label{invmet}G^{\alpha\beta} = l_{11}^2 V_3^{-2/3} L^{-2}\begin{pmatrix}
1 & A & 0  \\
A & A^2 +L^3/T_2 & L^3T_1/T_2 \\
0 & L^3T_1/T_2 &  L^3\vert T \vert^2/T_2
\end{pmatrix},\ee  
which on matching with \C{inverse}, using \C{newS1} and \C{newS2} gives us that
\bea \label{invert}
l_{11}^2 \hat\s &=& \frac{(1-A)V_3^{2/3}}{L^2}  ,\non \\ l_{11}^2 \hat\lambda &=& \Big[(1-T_1)\frac{L}{T_2} + \frac{A(A -1)}{L^2} \Big]V_3^{2/3}, \non \\ l_{11}^2 \hat\rho &=& (\vert T \vert^2 - T_1) \frac{LV_3^{2/3}}{T_2} ,\non \\ l_{11}^2 \hat\mu &=& \frac{AV_3^{2/3}}{L^2} , \non \\ l_{11}^2 \hat\theta &=& \frac{LT_1V_3^{2/3}}{T_2} , \eea
and
\be \label{invert2} {\hat\Delta}_3 = \frac{V_3^2}{l_{11}^6}.\ee
The Poisson resummed measure and the lattice sum satisfies the relation
\be \label{measure}d\hat\Upsilon  \sum_{\hat{k}^{\alpha I}} e^{-\pi^2 l_{11}^2 G_{IJ}G_{\alpha\beta} \hat{k}^{\alpha I} \hat{k}^{\beta J}} = \frac{4V_3^{7/3}}{l_{11}^{10}L^2 T_2^2} dV_3 dL dT_1 dT_2 dA \sum_{\hat{k}^{\alpha I}} e^{-\pi^2 \mathcal{V}_2 V_3^{2/3} \hat{G}_{IJ} \hat{G}_{\alpha\beta} \hat{k}^{\alpha I} \hat{k}^{\beta J}},\ee
where ${G}_{IJ} = \mathcal{V}_2 \hat{G}_{IJ}$ from \C{defmet}, and 
\be \label{needdef}
\hat{G}_{\alpha\beta} = \begin{pmatrix}
L^2 + A^2 \vert T  \vert^2/LT_2 & - A \vert T\vert^2/LT_2 & A T_1/LT_2  \\
- A \vert T\vert^2/LT_2 & \vert T\vert^2/LT_2 & -T_1/LT_2 \\
 A T_1/LT_2  & -T_1/LT_2& 1/LT_2
\end{pmatrix}.\ee 
Defining
\be d\mu \equiv \frac{1}{L^2 T_2^2} dL dT_1 dT_2 dA,\ee
we see that the unrenormalized amplitude for the $D^8\mathcal{R}^4$ interaction is given by
\be \label{mostimp}
I^{D^8\mathcal{R}^4} = \frac{\pi^{21/2}\Sigma_4}{64 l_{11}^{13}} \int_0^\infty d V_3 V_3^{10/3} \int d\mu f_1(L,T_1,T_2,A) \hat{F}_L ,\ee
where
\bea L^2 T_2^2 f_1(L,T_1,T_2,A) &=& \Big[(1-A) (\vert T \vert^2 - T_1)^2 +A T_1^2 + T_1 (\vert T \vert^2 -T_1) \Big] \non \\ &&\times \Big[ (1-T_1)\frac{L}{T_2} + \frac{A(A-1)}{L^2 }\Big]\eea
and
\be \label{defpl}\hat{F}_L = \sum_{\hat{k}^{\alpha I}} e^{-\pi^2 \mathcal{V}_2 V_3^{2/3} \hat{G}_{IJ} \hat{G}_{\alpha\beta} \hat{k}^{\alpha I} \hat{k}^{\beta J}} .\ee
Similarly for the $D^{10} \mathcal{R}^4$ interaction we get that
\bea \label{mostimp2} I^{D^{10} \mathcal{R}^4} = \frac{\pi^{21/2}\Sigma_5}{768 l_{11}^{11}} \int_0^\infty d V_3 V_3^{8/3} \int d\mu f_2(L,T_1,T_2,A) \hat{F}_L ,\eea
where
\bea L T_2 f_2 (L,T_1,T_2,A)&=& \Big[ -1 + \frac{\vert T\vert^2}{2T_2} \Big( \frac{1}{L^3} + \frac{\vert T\vert^2}{T_2}\Big) + \frac{4\vert T \vert^2}{T_2} \Big( \frac{1-T_1}{T_2} +\frac{A(A-1)}{L^3}\Big) \non \\ &&-\frac{\vert T\vert^4}{T_2^2}\Big( \frac{1-T_1}{T_2} +\frac{A(A-1)}{L^3}\Big)^2 \Big] \Big[ (1-T_1)\frac{L}{T_2} + \frac{A(A-1)}{L^2 }\Big] \non \\ && \times\Big[ AT_1 + (1-A)(\vert T\vert^2 - T_1)\Big].\eea

In the above expressions \C{mostimp} and \C{mostimp2}, the integral over $V_3$ is explicit. However, the integral over the moduli $L,A,T_1,T_2$ is over an involved domain. To see this, from \C{invert} and \C{invert2}, we have that
\bea \label{map}&&T_1 = \frac{\hat\theta(\hat\s+\hat\mu)}{\hat\s\hat\mu +(\hat\lambda+\hat\theta)(\hat\s +\hat\mu)}, \quad T_2 = \frac{{\hat\Delta}_3^{1/2} \sqrt{\hat\s+\hat\mu}}{\hat\s\hat\mu +(\hat\lambda+\hat\theta)(\hat\s +\hat\mu)}, \non \\ &&A = \frac{\hat\mu}{\hat\s+\hat\mu},  \quad L= \frac{{\hat\Delta}_3^{1/6}}{\sqrt{\hat\s+\hat\mu}}, \quad V_3 = l_{11}^3 \hat{\Delta}_3^{1/2}\eea
leading to the constraints
\bea \label{constraint}
&&0 \leq   A, T_1 \leq 1, \quad T_2 \geq 0, \quad L \geq 0, \non \\&& \Big\vert T -\frac{1}{2}\Big\vert^2  = \frac{1}{4} + \frac{\hat\rho(\hat\s +\hat\mu)}{\hat\s\hat\mu +(\hat\lambda+\hat\theta)(\hat\s +\hat\mu)} \geq \frac{1}{4}, \non \\ &&\Big(A - \frac{1}{2}\Big)^2 + \frac{L^3}{T_2} = \frac{1}{4} + \frac{\hat\lambda+\hat\theta}{\hat\s+\hat\mu} \geq \frac{1}{4} , \non \\  &&\Big(A -\frac{1}{2}\Big)^2 +\frac{L^3}{T_2}\vert T-1\vert^2 =\frac{1}{4} + \frac{\hat\lambda+\hat\rho}{\hat\s+\hat\mu} \geq \frac{1}{4}.\eea
Thus the constraints \C{constraint} define the integral over the moduli $L,A,T_1,T_2$.

Note that the lattice sum $\hat{F}_L$ is an infinite sum over the KK winding mode integers that involves the inverse metric \C{invmet}. We shall find it very useful in our analysis to act with a particular $SL(3,\mathbb{Z})$ transformation $S$ on the inverse metric (and hence on the metric) that leaves $\hat{F}_L$ invariant. This sends the inverse metric $G^{-1} \rightarrow S^T G^{-1}S$. Consider the $SL(3,\mathbb{Z})$ transformation
\be S= \begin{pmatrix}
1 & -1 & 0  \\
0 & 1 & 0 \\
 0  & 0& 1
\end{pmatrix}.\ee
Under its action, 
\bea \label{equiF}G^{\alpha\beta} &&= l_{11}^2 V_3^{-2/3} L^{-2}\begin{pmatrix}
1 & A & 0  \\
A & A^2 +L^3/T_2 & L^3T_1/T_2 \\
0 & L^3T_1/T_2 &  L^3\vert T \vert^2/T_2
\end{pmatrix} \non \\ &&\rightarrow  l_{11}^2 V_3^{-2/3} L^{-2}\begin{pmatrix}
1 & A-1 & 0  \\
A-1 & (A-1)^2 +L^3/T_2 & L^3T_1/T_2 \\
0 & L^3T_1/T_2 &  L^3\vert T \vert^2/T_2 
\end{pmatrix} \non \\  &&\equiv l_{11}^2 V_3^{-2/3} L^{-2}\begin{pmatrix}
1 & 1-A & 0  \\
1-A & (1-A)^2 +L^3/T_2 & L^3T_1/T_2 \\
0 & L^3T_1/T_2 &  L^3\vert T \vert^2/T_2 
\end{pmatrix}.\eea 
The first two lines in \C{equiF} express the equivalence of the inverse metric under $S$. The last line stands for an equality only in the lattice sum, obtained by sending $\{l_I,m_I,n_I\} \rightarrow \{-l_I,m_I,n_I\}$. Thus this action simply interchanges $A$ with $1-A$. This corresponds to interchanging $\hat\s$ and $\hat\mu$ keeping $\hat\l,\hat\rho,\hat\theta$ fixed. Thus\footnote{The trivial $V_3$ dependence of $\hat{F}_L$ is always implicit.} 
\be \label{impdef}\hat{F}_L (L,T,\bar{T},A) = \hat{F}_L(L,T,\bar{T},1-A).\ee 
 
\section{The structure of the three loop $D^8\mathcal{R}^4$ amplitude}

We now consider in detail the structure of the unrenormalized three loop $D^8\mathcal{R}^4$  amplitude that results from the ladder diagrams given by \C{mostimp}. This integral has primitive three loop as well as one and two loop subdivergences in it, which have to be regularized. These ultraviolet divergences come from the regions in moduli space where the Schwinger parameters $\s,\lambda,\rho,\mu,\theta \rightarrow 0$. The degree of divergence depends on how many of these parameters vanish. Thus for the Poisson resummed Schwinger parameters $\hat\s,\hat\l,\hat\rho,\hat\mu,\hat\theta$ the ultraviolet divergences arise as they go to infinity. In terms of the auxiliary geometry, they arise from the boundaries of moduli space as $V_3, T_2,L\rightarrow \infty$\footnote{This simply follows from the structure of divergences of the Mercedes diagrams~\cite{Basu:2014hsa} which arise from the boundary of the fundamental domain of $SL(3,\mathbb{Z})$~\cite{Gordon,Grenier2,Grenier}. Our auxiliary geometry is a subspace of the $T^3$ geometry with the same boundary structure.}.  These divergences can be analyzed following~\cite{Basu:2014uba} where some of them have been explicitly calculated. This is done by directly calculating the one, two and three loop primitive divergences by looking at the structure of the loop diagrams. Then the divergences are regularized by adding counterterms by making use of results in the type II theory. The calculations are considerably simpler than the three loop amplitude itself because they only involve one and two loop supergravity amplitudes along with insertions of counterterm vertices\footnote{The three loop primitive divergences for the $D^8\mathcal{R}^4$ and $D^{10} \mathcal{R}^4$ interactions which are $O(\Lambda^{13})$ and $O(\Lambda^{11})$ respectively cancel as the structure is inconsistent with type II string theory~\cite{Basu:2014uba}. Here $\Lambda$ is the ultraviolet cutoff.}. Thus the divergent contributions to these amplitudes can be regularized to yield finite answers. 

It is easy to see the structure of the leading three loop primitive divergence that arises for the $D^8\mathcal{R}^4$ and $D^{10}\mathcal{R}^4$ interactions. This is simply obtained as $V_3 \rightarrow \infty$ with all the lattice momenta set to zero, so that the ultraviolet divergences are not regularized at the boundary of moduli space. Thus from \C{mostimp}, the divergence $\sim V_3^{13/3} l_{11}^{-13} \sim \hat\Delta_3^{13/6} \sim \Lambda^{13}$ for the $D^8\mathcal{R}^4$ interaction.
Similarly, from \C{mostimp2} we see that the three loop primitive divergence $\sim \Lambda^{11}$ for the $D^{10}\mathcal{R}^4$ interaction.

What about the finite contributions that do not require renormalization? These have to be obtained by calculating the three loop amplitude directly with appropriately chosen non--vanishing winding momenta that render the integrals finite at the boundaries of moduli space. We now consider in detail the structure of this amplitude.

From \C{constraint}, we note that $T$ satisfies the constraint
\be 0 \leq T_1 \leq 1, \quad T_2 \geq 0, \quad \Big\vert T- \frac{1}{2} \Big\vert^2 \geq \frac{1}{4},\ee 
which is the region $\mathcal{R}_T = f_1 \oplus f_2 \oplus g_1 \oplus g_2 \oplus h_1 \oplus h_2$ in figure 6.

\begin{figure}[ht]
\begin{center}
\[
\mbox{\begin{picture}(150,90)(0,0)
\includegraphics[scale=.55]{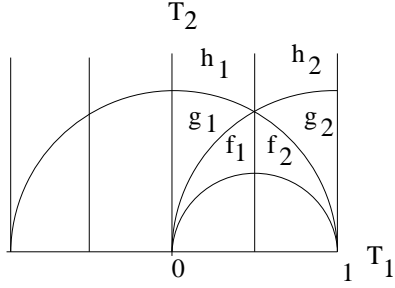}
\end{picture}}
\]
\caption{The $T$ plane}
\end{center}
\end{figure}

 We shall now map these 6 regions using appropriate $SL(2,\mathbb{Z})_T$ transformations into parts of the fundamental domain $\mathcal{F}_T$ of $SL(2,\mathbb{Z})_T$ denoted by $F$ and $F'$ in figure 7,~\footnote{Note that $\mathcal{F}_T = F \oplus F'$ is defined by
\be -1/2 \leq T_1 \leq 1/2, \quad T_2 \geq 0, \quad \vert T \vert^2 \geq 1.\ee} along the lines of~\cite{Green:1999pu,Green:2005ba}. These shall be accompanied by suitable transformations for $A$ and $L$ as described below.

\begin{figure}[ht]
\begin{center}
\[
\mbox{\begin{picture}(150,90)(0,0)
\includegraphics[scale=.55]{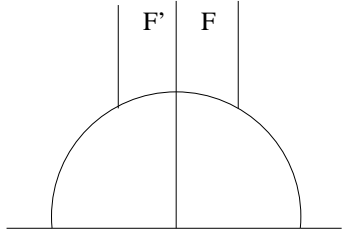}
\end{picture}}
\]
\caption{$\mathcal{F} = F \oplus F'$}
\end{center}
\end{figure}
 
We state the contribution   
\be \label{breakcont} \int d\mu f_1(L,T_1,T_2,A) \hat{F}_L \ee
to the $D^8\mathcal{R}^4$ term from each patch. 

{\bf{(i)}} The region $h_1$ is trivially mapped to $F$ using the identity map. We set
\be T\rightarrow T' = T, \quad A \rightarrow A' = A, \quad L\rightarrow L' = L.\ee
Thus the contribution of \C{breakcont} from the region $h_1$, 
along with the constraints \C{constraint} gives us 
\be \label{h1} \int_F d\mu f_1^{(1)}(L,T_1,T_2,A) \hat{F}_L (L,T,A)\ee
along with
\be 0 \leq A \leq 1, \quad \Big(A - \frac{1}{2}\Big)^2 + \frac{L^3}{T_2} \geq \frac{1}{4}, \quad \Big(A - \frac{1}{2}\Big)^2 + \frac{L^3}{T_2}\vert T-1\vert^2 \geq  \frac{1}{4},\ee
and
\bea f_1^{(1)}(L,T_1,T_2,A) &=& \Big[(1-A) (\vert T \vert^2 - T_1)^2 +A T_1^2 + T_1 (\vert T \vert^2 -T_1) \Big] \non \\ &&\times \frac{1}{L^2 T_2^2}\Big[ (1-T_1)\frac{L}{T_2} + \frac{A(A-1)}{L^2 }\Big].\eea
The lattice sum $\hat{F}_L (L,T,A)$ is given by \C{defpl} where $T$ is the complex structure of the $SL(2,\mathbb{Z})_T$ torus. Hence the inverse metric $G^{\alpha\beta}$ which enters the lattice sum before Poisson resummation is given by \C{invmet}, and thus $\hat{G}_{\alpha\beta}$ is given by \C{needdef}.

{\bf{(ii)}} The region $f_2$ is mapped to to $F'$. We set
\be T \rightarrow T' = T/(1-T), \quad A\rightarrow A' = 1-A, \quad L' \rightarrow L.\ee
Thus we get that
\bea \label{need1}
&& l_{11}^2 \hat\s = \frac{A'V_3^{2/3}}{L'^2}  , \quad l_{11}^2 \hat\lambda = \Big[(1+T'_1)\frac{L'}{T'_2} + \frac{A'(A' -1)}{L'^2} \Big]V_3^{2/3}, \non \\ && l_{11}^2 \hat\rho = -\frac{T'_1 L'V_3^{2/3}}{T'_2} , \quad l_{11}^2 \hat\mu = \frac{(1-A')V_3^{2/3}}{L'^2} , \quad l_{11}^2 \hat\theta = (\vert T'\vert^2 + T'_1)\frac{L'V_3^{2/3}}{T'_2} . \eea
Thus the contribution of \C{breakcont} from the region $f_2$, 
along with the constraints \C{constraint} gives us 
\be \label{f2} \int_{F'} d\mu f_1^{(1)}(L,-T_1,T_2,A) \hat{F}_L (L,T,A)\ee
along with
\be  0 \leq A \leq 1, \quad \Big(A - \frac{1}{2}\Big)^2 + \frac{L^3}{T_2} \geq \frac{1}{4}, \quad \Big(A - \frac{1}{2}\Big)^2 + \frac{L^3}{T_2}\vert T +1\vert^2 \geq  \frac{1}{4}.\ee
To see that the lattice sums $\hat{F}_L$ in \C{h1} and \C{f2} are equal, we note that from \C{invert} and \C{need1} it follows that they are related by the transformations $\hat\s \leftrightarrow \hat\mu, \hat\theta \leftrightarrow \hat\rho$ keeping $\hat\lambda$ fixed (including the change of sign of $T_1$). This is precisely the $P_3$ transformation of \C{listtrans} which is a symmetry of the ladder skeleton. Hence the lattice sum in $F_L$ in \C{f2} involves $G^{\alpha\beta}$ in \C{invmet} with $T_1 \rightarrow - T_1$, which is absorbed by sending $n_I \rightarrow - n_I$ in the lattice sum.

Thus the contributions from $h_1 \oplus f_2$ add to give
\be \label{regone} \int_{\mathcal{F}_T} d\mu f_1^{(1)}(L,\vert T_1 \vert,T_2,A) \hat{F}_L (L, T,A)\ee
along with the constraints
\be  \label{contone} 0 \leq A \leq 1, \quad \Big(A - \frac{1}{2}\Big)^2 + \frac{L^3}{T_2} \geq \frac{1}{4}, \quad \Big(A - \frac{1}{2}\Big)^2 + \frac{L^3}{T_2}\Big( (\vert T_1 \vert -1)^2 + T_2^2 \Big) \geq  \frac{1}{4}.\ee

{\bf{(iii)}} The region $f_1$ is mapped to $F$. We set
\be T \rightarrow T' = (T-1)/T, \quad A \rightarrow A'=A, \quad L \rightarrow L' =L.\ee
Thus
\bea \label{need2}
&& l_{11}^2 \hat\s = \frac{(1-A')V_3^{2/3}}{L'^2}  , \quad l_{11}^2 \hat\lambda = \Big[(\vert T'\vert^2 -T'_1)\frac{L'}{T'_2} + \frac{A'(A' -1)}{L'^2} \Big]V_3^{2/3}, \non \\ && l_{11}^2 \hat\rho = \frac{T'_1 L'V_3^{2/3}}{T'_2} , \quad l_{11}^2 \hat\mu = \frac{A'V_3^{2/3}}{L'^2} , \quad l_{11}^2 \hat\theta = (1- T'_1)\frac{L'V_3^{2/3}}{T'_2} . \eea
Thus the contribution of \C{breakcont} from the region $f_1$, 
along with the constraints \C{constraint} gives us 
\be \label{f1} \int_{F} d\mu f_1^{(2)}(L,T_1,T_2,A) \hat{F}_L (L,1/(1-T),A)\ee
along with
\be  0 \leq A \leq 1, \quad \Big(A - \frac{1}{2}\Big)^2 + \frac{L^3}{T_2} \vert T \vert^2 \geq \frac{1}{4}, \quad \Big(A - \frac{1}{2}\Big)^2 + \frac{L^3}{T_2}\vert T -1\vert^2 \geq  \frac{1}{4},\ee
and
\bea f_1^{(2)} (L,T_1,T_2,A) &=& \Big[(1-A) T_1^2 +A(1-T_1)^2 + T_1 (1 -T_1) \Big] \non \\ &&\times \frac{1}{L^2 T_2^2}\Big[ (\vert T \vert^2 -T_1)\frac{L}{T_2} + \frac{A(A-1)}{L^2 }\Big].\eea
Hence the lattice sum $\hat{F}_L (L,1/(1-T),A)$ involves the metric
\be \label{needdef2}
\hat{G}_{\alpha\beta} = \begin{pmatrix}
L^2 + A^2 /LT_2 & - A /LT_2 & A (1-T_1)/LT_2  \\
- A /LT_2 & 1/LT_2 & -(1-T_1)/LT_2 \\
 A (1-T_1)/LT_2  & -(1-T_1)/LT_2& \vert 1-T\vert^2/LT_2
\end{pmatrix}.\ee 
Thus before Poisson resumming, the lattice sum $F_L$ involves the inverse metric
\be \label{invmet2}G^{\alpha\beta} = l_{11}^2 V_3^{-2/3} L^{-2}\begin{pmatrix}
1 & A & 0  \\
A & A^2 +L^3\vert 1-T\vert^2/T_2 & L^3(1-T_1)/T_2 \\
0 & L^3(1-T_1)/T_2 &  L^3/T_2
\end{pmatrix}.\ee  
As discussed before, the lattice sum is invariant under any transformation $G^{-1} \rightarrow S^T G^{-1} S$, where $S$ is a matrix with integer entries. Taking $S$ to be the $SL(3,\mathbb{Z})$ matrix
\be S = \begin{pmatrix}
1 & 0 & 0  \\
0 & 0 & -1 \\
0 & 1 &  1
\end{pmatrix},\ee 
we see that we can equivalently use the inverse metric
\be \label{mat}G^{\alpha\beta} = l_{11}^2 V_3^{-2/3} L^{-2}\begin{pmatrix}
1 & 0 & -A  \\
0 & L^3/T_2 & L^3 T_1/T_2 \\
-A & L^3 T_1/T_2 &  A^2 + L^3\vert T \vert^2/T_2
\end{pmatrix}.\ee
Note that if we send $T_1 \rightarrow -T_1$ in \C{mat}, $F_L$ (and hence $\hat{F}_L$) remains invariant simply by redefining $m_I \rightarrow -m_I$.

{\bf{(iv)}} The region $g_1$ is mapped to to $F'$. We set
\be T \rightarrow T' = -1/T, \quad A \rightarrow A' = 1-A, \quad L \rightarrow L' =L.\ee
Thus
\bea \label{need3}
&& l_{11}^2 \hat\s = \frac{A'V_3^{2/3}}{L'^2}  , \quad l_{11}^2 \hat\lambda = \Big[(\vert T'\vert^2 +T'_1)\frac{L'}{T'_2} + \frac{A'(A' -1)}{L'^2} \Big]V_3^{2/3}, \non \\ && l_{11}^2 \hat\rho = \frac{(1+T'_1) L'V_3^{2/3}}{T'_2} , \quad l_{11}^2 \hat\mu = \frac{(1-A')V_3^{2/3}}{L'^2} , \quad l_{11}^2 \hat\theta = -\frac{T_1' L'V_3^{2/3}}{T'_2} . \eea
Thus the contribution of \C{breakcont} from the region $g_1$, 
along with the constraints \C{constraint} gives us 
\be \label{g1} \int_{F'} d\mu f_1^{(2)}(L,-T_1,T_2,A) \hat{F}_L (L,1/(1-T),A)\ee
along with
\be  0 \leq A \leq 1, \quad \Big(A - \frac{1}{2}\Big)^2 + \frac{L^3}{T_2} \vert T \vert^2 \geq \frac{1}{4}, \quad \Big(A - \frac{1}{2}\Big)^2 + \frac{L^3}{T_2}\vert T + 1\vert^2 \geq  \frac{1}{4}.\ee
To see that the lattice sums $\hat{F}_L$ in \C{f1} and \C{g1} are equal, we note that from \C{need2} and \C{need3} it follows that they are related by the transformations $\hat\s \leftrightarrow \hat\mu, \hat\theta \leftrightarrow \hat\rho$ keeping $\hat\lambda$ fixed (including the change of sign of $T_1$), which is the $P_3$ transformation. Thus the two lattice sums follow from \C{mat} and are the same as discussed above.

Thus the contributions from $f_1 \oplus g_1$ add to give
\be \label{regtwo} \int_{\mathcal{F}_T} d\mu f_1^{(2)} (L,\vert T_1 \vert,T_2,A) \hat{F}_L (L, 1/(1-T),A)\ee
along with the constraints
\be  \label{conttwo} 0 \leq A \leq 1, \quad \Big(A - \frac{1}{2}\Big)^2 + \frac{L^3}{T_2} \vert T \vert^2 \geq \frac{1}{4}, \quad \Big(A - \frac{1}{2}\Big)^2 + \frac{L^3}{T_2}\Big( (\vert T_1 \vert -1)^2 + T_2^2 \Big) \geq  \frac{1}{4}.\ee

{\bf{(v)}} The region $g_2$ is mapped to $F$. We set
\be T \rightarrow T' = 1/(1-T), \quad  A \rightarrow A' = A, \quad L \rightarrow L' =L.\ee
Thus
\bea \label{need4}
&& l_{11}^2 \hat\s = \frac{(1-A')V_3^{2/3}}{L'^2}  , \quad l_{11}^2 \hat\lambda = \Big[\frac{T_1' L'}{T'_2} + \frac{A'(A' -1)}{L'^2} \Big]V_3^{2/3}, \non \\ && l_{11}^2 \hat\rho = \frac{(1-T'_1) L'V_3^{2/3}}{T'_2} , \quad l_{11}^2 \hat\mu = \frac{A'V_3^{2/3}}{L'^2} , \quad l_{11}^2 \hat\theta = (\vert T' \vert^2-T_1')\frac{ L'V_3^{2/3}}{T'_2} . \eea
Thus the contribution of \C{breakcont} from the region $g_2$, 
along with the constraints \C{constraint} gives us 
\be \label{g2} \int_{F} d\mu f_1^{(3)} (L,T_1,T_2,A) \hat{F}_L (L,(T-1)/T,A)\ee
along with
\be  0 \leq A \leq 1, \quad \Big(A - \frac{1}{2}\Big)^2 + \frac{L^3}{T_2} \geq \frac{1}{4}, \quad \Big(A - \frac{1}{2}\Big)^2 + \frac{L^3}{T_2}\vert T\vert^2 \geq  \frac{1}{4},\ee 
and
\bea f_1^{(3)} (L,T_1,T_2,A) &=& \Big[(1-A) (1-T_1)^2 +A (\vert T \vert^2 -T_1)^2 + (1-T_1) (\vert T \vert^2 -T_1) \Big]  \non \\ && \times \frac{1}{L^2 T_2^2} \Big[ \frac{T_1 L}{T_2} + \frac{A(A-1)}{L^2 }\Big].\eea
Hence the lattice sum $\hat{F}_L (L,(T-1)/T,A)$ involves the metric
\be \label{needdef3}
\hat{G}_{\alpha\beta} = \begin{pmatrix}
L^2 + A^2 \vert 1-T\vert^2/LT_2 & - A \vert 1-T\vert^2 /LT_2 & A (\vert T\vert^2-T_1)/LT_2  \\
- A \vert 1-T\vert^2/LT_2 & \vert 1-T\vert^2/LT_2 & -(\vert T\vert^2-T_1)/LT_2 \\
 A (\vert T\vert^2-T_1)/LT_2  & -(\vert T \vert^2-T_1)/LT_2& \vert T\vert^2/LT_2
\end{pmatrix}.\ee 
Thus before Poisson resumming, the lattice sum $F_L$ involves the inverse metric
\be \label{invmet3} G^{\alpha\beta} = l_{11}^2 V_3^{-2/3} L^{-2}\begin{pmatrix}
1 & A & 0  \\
A & A^2 +L^3\vert T\vert^2/T_2 & L^3(\vert T \vert^2-T_1)/T_2 \\
0 & L^3(\vert T \vert^2-T_1)/T_2 &  L^3\vert T-1\vert^2/T_2
\end{pmatrix}.\ee  
Making the transformation $G^{-1} \rightarrow S^T G^{-1} S$, where $S$ is the $SL(3,\mathbb{Z})$ matrix 
\be S = \begin{pmatrix}
1 & 0 & 0  \\
0 & 1 & 1 \\
0 & -1 &  0
\end{pmatrix},\ee 
we see that we can equivalently use the inverse metric
\be G^{\alpha\beta} = l_{11}^2 V_3^{-2/3} L^{-2}\begin{pmatrix}
1 & A & A  \\
A & A^2 + L^3/T_2 & A^2 + L^3 T_1/T_2 \\
A & A^2 + L^3 T_1/T_2 &  A^2 + L^3\vert T \vert^2/T_2
\end{pmatrix}.\ee
Unlike the earlier cases, we see that now $F_L$ is not invariant under $T_1 \rightarrow -T_1$
. For later convenience, we denote
\be \hat{F}_L (L,(T-1)/T,A)\equiv {\hat{\mathcal{F}}}_L (L,T_1,T_2,A).\ee

{\bf{(vi)}} The region $h_2$ is mapped to $F'$. We set
\be T \rightarrow T' = T-1, \quad  A \rightarrow A' = 1-A, \quad L \rightarrow L' =L.\ee
Thus
\bea \label{need5}
&& l_{11}^2 \hat\s = \frac{A'V_3^{2/3}}{L'^2}  , \quad l_{11}^2 \hat\lambda = \Big[-\frac{T_1' L'}{T'_2} + \frac{A'(A' -1)}{L'^2} \Big]V_3^{2/3}, \non \\ && l_{11}^2 \hat\rho = \frac{(\vert T\vert^2 + T'_1) L'V_3^{2/3}}{T'_2} , \quad l_{11}^2 \hat\mu = \frac{(1-A')V_3^{2/3}}{L'^2} , \quad l_{11}^2 \hat\theta = (1+T_1')\frac{ L'V_3^{2/3}}{T'_2} . \eea
Thus the contribution of \C{breakcont} from the region $h_2$, 
along with the constraints \C{constraint} gives us 
\be \label{h2} \int_{F'} d\mu f_1^{(3)} (L,-T_1,T_2,A) {\hat{\mathcal{F}}}_L (L,-T_1,T_2,A)\ee
along with
\be  0 \leq A \leq 1, \quad \Big(A - \frac{1}{2}\Big)^2 + \frac{L^3}{T_2} \geq \frac{1}{4}, \quad \Big(A - \frac{1}{2}\Big)^2 + \frac{L^3}{T_2}\vert T\vert^2 \geq  \frac{1}{4}.\ee
To see that the lattice sums $\mathcal{F}_L$ in \C{g2} and \C{h2} are equal upto the opposite sign for $T_1$, we note that from \C{need4} and \C{need5} it follows that they are related by the transformations $\hat\s \leftrightarrow \hat\mu, \hat\theta \leftrightarrow \hat\rho$ keeping $\hat\lambda$ fixed (upto the change of sign of $T_1$), which is the $P_3$ transformation. 

Thus the contributions from $g_2 \oplus h_2$ add to give
\be \label{regthree} \int_{\mathcal{F}_T} d\mu f_1^{(3)}(L,\vert T_1 \vert,T_2,A) {\hat{\mathcal{F}}}_L (L, \vert T_1 \vert, T_2 ,A)\ee
along with the constraints
\be  \label{contthree} 0 \leq A \leq 1, \quad \Big(A - \frac{1}{2}\Big)^2 + \frac{L^3}{T_2} \geq \frac{1}{4}, \quad \Big(A - \frac{1}{2}\Big)^2 + \frac{L^3}{T_2}\vert T  \vert^2 \geq  \frac{1}{4}.\ee

We can further simplify the contributions given in \C{regone}, \C{regtwo} and \C{regthree} by using \C{impdef}. For example, we write \C{regone} as
\be \frac{1}{2} \int_{\mathcal{F}_T} d\mu \Big[ f_1^{(1)} (L,\vert T_1 \vert, T_2,A) + f_1^{(1)} (L,\vert T_1 \vert, T_2, 1-A)\Big] \hat{F}_L (L,T,A)\ee
where we have substituted $A\rightarrow 1 - A$ in the second term and used \C{impdef}, as $d\mu$ and \C{contone} are invariant under $A\rightarrow 1-A$.
Thus the total contribution is given by
\be C_1 + C_2 +C_3,\ee
where

{\bf{(i)}} \be \label{C1}C_1 = \int_{\mathcal{F}_T} d\mu F_1 (L,\vert T_1 \vert ,T_2,A) \hat{F}_L (L,T,A)\ee
where
\be F_1 (L,\vert T_1 \vert ,T_2,A) = \frac{\vert T\vert^4}{2L^2 T_2^2} \Big[ (1-\vert T_1 \vert)\frac{L}{T_2} + \frac{A(A-1)}{L^2 }\Big]\ee
along with the constraints \C{contone},

{\bf{(ii)}} \be \label{C2} C_2 = \int_{\mathcal{F}_T} d\mu F_2 (L,\vert T_1 \vert ,T_2,A) \hat{F}_L (L,1/(1-T),A)\ee
where
\be F_2 (L,\vert T_1 \vert ,T_2,A) = \frac{1}{2L^2 T_2^2} \Big[ (\vert T\vert^2 -\vert T_1\vert)\frac{L}{T_2} + \frac{A(A-1)}{L^2 }\Big]\ee
along with the constraints \C{conttwo}, and

{\bf{(iii)}} \be \label{C3}C_3 = \int_{\mathcal{F}_T} d\mu F_3 (L,\vert T_1 \vert ,T_2,A) {\hat{\mathcal{F}}}_L (L, \vert T_1 \vert, T_2 ,A) \ee
where
\be F_3 (L,\vert T_1 \vert ,T_2,A) = \frac{1}{2L^2 T_2^2} (\vert T\vert^2 -2 \vert T_1\vert+1)^2\Big[ \frac{ \vert T_1 \vert L}{T_2} + \frac{A(A-1)}{L^2 }\Big]\ee
along with the constraints \C{contthree}.

Thus the unrenormalized $D^8\mathcal{R}^4$ interaction is given by
\be \label{impmost3}
I^{D^8\mathcal{R}^4} = \frac{\pi^{21/2}\Sigma_4}{64 l_{11}^{13}} \int_0^\infty d V_3 V_3^{10/3} (C_1 + C_2 + C_3) .\ee

In order to evaluate \C{impmost3}, one possible strategy would be to evaluate $\Delta_{\Omega} I^{D^8\mathcal{R}^4}$, where $\Delta_\Omega$ is the $SL(2,\mathbb{Z})$ invariant Laplacian
\be \Delta_\Omega = 4\Omega_2^2\frac{\p^2}{\p\Omega \p\bar\Omega}.\ee  
This operator acts only on the lattice factor in \C{impmost3}, and one can hope to replace this action on $\hat{F}_L$ with essentially the action of the Laplacian on the moduli space of the auxiliary geometry. This happens at two loops~\cite{Green:1999pu,Green:2005ba} and for the Mercedes diagrams at three loops~\cite{Basu:2014hsa}, where the auxiliary geometries are $T^2$ and $T^3$ respectively. However, we do not have a detailed understanding of the auxiliary geometry in our case\footnote{Attempts to construct such Laplacians always seem to leave behind terms that involve lattice momenta in the integrand beyond the exponential factor. These contributions have to be separately calculated.}. Thus we shall restrict ourselves to contributions to the amplitude for specific values of the moduli. 
Since we have to integrate over the moduli space, our results give us exact expressions for certain values of the parameters in the integrals\footnote{The moduli we choose lie between 0 and 1, hence one can make a perturbative expansion in the moduli around these values to get the complete answer.}. 
It would be very interesting to perform the integrals over the moduli space to get the exact answer.          

What are the specific values of the moduli we choose? These are values for which the integrals simplify and can be done exactly. There is a justification based on the structure of supersymmetry about the role played by precisely these choices of moduli in determining the various interactions, which we discuss in appendix E. 

We now consider these contributions in detail. 

\subsection{$A=0 + A=1$ in \C{mostimp}}

We now calculate the exact contribution from $A=0$ and $A=1$ together in \C{mostimp}. Thus this includes the total contribution coming from the lower and upper ends of the $A$ integration in \C{constraint}. In the $\Theta$ plane given in figure 11, this corresponds to the lines $\Theta_1 =0$ and $\Theta_1 =1$.

To begin with, consider the case when $A=0$ in \C{impmost3}. In terms of the original variables, this contains contributions from $A=0$ as well as $A=1$ dictated by the maps described above. When $A=0$, the lattice factors in \C{C1}, \C{C2} and \C{C3} become $\hat{F}_L (L,T,0)$, $\hat{F}_L (L,1/(1-T),0)$ and ${\hat{\mathcal{F}}}_L (L, \vert T_1 \vert, T_2 ,0)$ respectively. Now for any $T$, $\hat{F}_L (L,T,0)$ involves the block diagonal inverse metric
\bea G^{\alpha\beta} = l_{11}^2 V_3^{-2/3} L^{-2}\begin{pmatrix}
1 & 0 & 0  \\
0 & L^3/T_2 & L^3T_1/T_2 \\
0 & L^3T_1/T_2 &  L^3\vert T \vert^2/T_2
\end{pmatrix} \eea
which has an $SL(2,\mathbb{Z})_T$ invariant subspace. This allows us to perform the integrals exactly. Also $\hat{F}_L (L,T,0) = \hat{F}_L (L,1/(1-T),0)$ by performing the $SL(2,\mathbb{Z})_T$ transformation $T\rightarrow 1/(1-T)$ which is implemented by the $SL(3,\mathbb{Z})$ matrix 
\be S= \begin{pmatrix}
1 & 0 & 0  \\
0 & 1 & 1 \\
0 & -1 &  0
\end{pmatrix}. \ee
Also ${\hat{\mathcal{F}}}_L (L, \vert T_1 \vert, T_2 ,0)$ involves the inverse metric
\bea G^{\alpha\beta} = l_{11}^2 V_3^{-2/3} L^{-2}\begin{pmatrix}
1 & 0 & 0  \\
0 & L^3/T_2 & L^3 \vert T_1\vert /T_2 \\
0 & L^3 \vert T_1 \vert/T_2 &  L^3\vert T \vert^2/T_2
\end{pmatrix} \eea
and hence is independent of the sign of $T_1$ which follows by sending $m_I \rightarrow -m_I$. Hence
\be \hat{F}_L (L,T,0) = \hat{F}_L (L,1/(1-T),0) = {\hat{\mathcal{F}}}_L (L, \vert T_1 \vert, T_2 ,0) \equiv \hat{F}_L (L,T).\ee
Note that we have that $0 \leq A \leq 1$, and hence this is the lower boundary of the integral. This is not a boundary of moduli space and hence is not associated with ultraviolet divergences\footnote{The nested divergences that arise in the integrals arise from the boundary of moduli space.}. Thus we see that when $A=0$ in \C{impmost3}
\bea \label{A0}
I^{D^8\mathcal{R}^4} = \frac{\pi^{21/2}\Sigma_4}{128 l_{11}^{13}} \int_0^\infty d V_3 V_3^{10/3} \int_0^\infty \frac{dL}{L^2} \int_{\mathcal{F}_T} \frac{dT_1 dT_2}{T_2^2}  F^{D^8\mathcal{R}^4} (T,\bar{T}) \frac{\hat{F}_L (L,T)}{L},\eea
where
\be \label{FD8R4}F^{D^8\mathcal{R}^4} (T,\bar{T})= \frac{1}{T_2^3} \Big[ \vert T\vert^4 (1-\vert T_1 \vert) + \vert T \vert^2 - \vert T_1 \vert +\vert T_1 \vert (\vert T\vert^2 -2\vert T_1 \vert +1)^2\Big].\ee 

Exactly the same is the analysis when $A=1$ in \C{impmost3}, which also receives contributions from both $A=0$ and $A=1$ in \C{mostimp}. In fact, the total contribution from $A=0$ and $A=1$ in \C{mostimp} is simply twice the contribution from \C{A0}.  

We now evaluate the finite part of \C{A0}. As discussed before, the primitive three loop divergent contribution is of the form $\Lambda^{13}$. Thus on dimensional grounds, we get that
\be I^{D^8\mathcal{R}^4} = a_0 \Lambda^{13} +\ldots + (l_{11}^2 \mathcal{V}_2)^{-13/2} f(\Omega,\bar\Omega),\ee
where $a_0$ is a moduli independent constant and $f(\Omega,\bar\Omega)$ is the $SL(2,\mathbb{Z})$ invariant finite part of $I^{D^8\mathcal{R}^4}$. The remaining contributions involve various one loop and two loop divergences which can also be evaluated along the lines of our analysis.

In the expression for $\hat{F}_L$ the terms involving $\hat{l}^I$ separate from the terms involving $\hat{m}^I, \hat{n}^I$, hence radically simplifying our analysis. The $V_3$ and $L$ dependence of these two contributions are $V_3^{2/3} L^2$ and $V_3^{2/3} L^{-1}$ in the exponentials respectively. Hence defining
\be V_3^{2/3} L^2 = x, \quad V_3^{2/3} L^{-1} = y,\ee
the $x$ integral which involves $\hat{l}^I$ can be easily done to yield
\bea \label{A01}
 I^{D^8\mathcal{R}^4} &=& \frac{\pi^8E_{3/2} (\Omega,\bar\Omega)\Sigma_4 }{512 (l_{11}^2 \mathcal{V}_2)^{13/2}} \int_0^\infty d y y^4 \int_{\mathcal{F}_T} \frac{dT_1 dT_2}{T_2^2} F^{D^8\mathcal{R}^4} (T,\bar{T}) \sum_{\hat{m}^I,\hat{n}^I} e^{-\pi^2 y \hat{G}_{IJ} (\hat{m} +\hat{n}T)^I(\hat{m}+\hat{n}\bar{T})^J/T_2}    \non \\ &\equiv& \frac{\pi^8E_{3/2} (\Omega,\bar\Omega)\Sigma_4 }{512 (l_{11}^2 \mathcal{V}_2)^{13/2}} \mathcal{I}^{D^8\mathcal{R}^4} (\Omega,\bar\Omega). \eea

In \C{A01}, we have used the expression for the $SL(2,\mathbb{Z})$ invariant non--holomorphic Eisenstein series\footnote{$E_s$ satisfies the Laplace equation
\be 4\Omega_2^2 \frac{\p^2 E_s}{\p\Omega \p\bar\Omega} = s(s-1) E_s.\ee}
\bea \label{Eisenstein} && E_s(\Omega,\bar\Omega) = \sum_{l_i \in \mathbb{Z}, (l_1,l_2) \neq (0,0)} \frac{\Omega_2^s}{\vert l_1 + l_2 \Omega\vert^{2s}} \non \\ && = 2\zeta(2s) \Omega_2^s + 2\sqrt{\pi} \Omega_2^{1-s} \frac{\Gamma(s-1/2)}{\Gamma(s)}\zeta(2s-1)\non \\ && +\frac{4\pi^s \sqrt{\Omega_2}}{\Gamma(s)} \sum_{k\in \mathbb{Z}, k\neq 0} \vert k \vert^{s-1/2} \mu(\vert k\vert, s) K_{s-1/2} (2\pi \Omega_2 \vert k \vert) e^{2\pi i k\Omega_1},\eea
where
\be \mu (k,s) = \sum_{m>0,m|k} \frac{1}{m^{2s-1}}.\ee

Note that in the $x$ integral we have kept only the ultraviolet finite part which yields the Eisenstein series\footnote{The divergent part comes when $\hat{l}^I =0$.}. The remaining part of the expression in \C{A01} involves an $SL(2,\mathbb{Z})_T$ invariant lattice sum, which is divergent from which we have to extract the finite part. 

To obtain the finite part of $\mathcal{I}^{D^8\mathcal{R}^4}$ in \C{A01}, consider
\bea \Delta_\Omega \mathcal{I}^{D^8\mathcal{R}^4} = \int_0^\infty d y y^4 \int_{\mathcal{F}_T} \frac{dT_1 dT_2}{T_2^2} F^{D^8\mathcal{R}^4} (T,\bar{T}) \Delta_\Omega\sum_{\hat{m}^I,\hat{n}^I} e^{-\pi^2 y \hat{G}_{IJ} (\hat{m} +\hat{n}T)^I(\hat{m}+\hat{n}\bar{T})^J/T_2} \non \\  = \int_0^\infty d y y^4 \int_{\mathcal{F}_T} \frac{dT_1 dT_2}{T_2^2} F^{D^8\mathcal{R}^4} (T,\bar{T}) \Delta_T \sum_{\hat{m}^I,\hat{n}^I} e^{-\pi^2 y \hat{G}_{IJ} (\hat{m} +\hat{n}T)^I(\hat{m}+\hat{n}\bar{T})^J/T_2}. \eea
We now integrate by parts so that $\Delta_T$ acts not on the lattice sum but on $F^{D^8\mathcal{R}^4}$. In doing so, we ignore boundary contributions which yield ultraviolet divergences, along the lines of~\cite{Green:1999pu,Green:2005ba}. Also in the lattice sum, we set $(\hat{m}^1,\hat{m}^2)\neq(0,0), (\hat{n}^1,\hat{n}^2)\neq (0,0)$ to get the finite contribution. Thus
\be \Delta_\Omega \mathcal{I}^{D^8\mathcal{R}^4}_{finite}   = \int_0^\infty d y y^4 \int_{\mathcal{F}_T} \frac{dT_1 dT_2}{T_2^2} \Delta_T F^{D^8\mathcal{R}^4} (T,\bar{T}) \sum_{{}^{(\hat{m}^1,\hat{m}^2)\neq(0,0)}_{(\hat{n}^1,\hat{n}^2)\neq (0,0)}} e^{-\pi^2 y \hat{G}_{IJ} (\hat{m} +\hat{n}T)^I(\hat{m}+\hat{n}\bar{T})^J/T_2}.\ee
Thus we need to analyze the structure of $\Delta_T F^{D^8\mathcal{R}^4}$. To do so, it is very useful to define
\be \label{calT} \mathcal{T}= \vert T_1 \vert - T_1^2 .\ee
We see that $F^{D^8\mathcal{R}^4}$ splits into a sum of functions each of which satisfies a Poisson equation. The structure of these equations is obtained recursively~\cite{Green:2008bf}. We start with the leading term in the small $T_2$ limit, and construct a Poisson equation which has it as the leading term in this limit. Subtracting this contribution from $F^{D^8 \mathcal{R}^4}$, we consider again the leading term in the small $T_2$ limit, and again construct a Poisson equation. This iteration goes on till all the terms in $F^{D^8 \mathcal{R}^4}$ are exhausted. To do this, we note that
\be  F^{D^8\mathcal{R}^4} (T,\bar{T})= -\frac{3\mathcal{T}^2}{T_2^3}+ \frac{1+2\mathcal{T}}{T_2} +T_2.\ee
This leads to
\bea F^{D^8\mathcal{R}^4} (T,\bar{T})= b_1^{(8)} (T,\bar{T})+ b_2^{(8)} (T,\bar{T}),\eea
where $b_1^{(8)}$ and $b_2^{(8)}$ are given by
\bea b_1^{(8)} &=& -\frac{3\mathcal{T}^2}{T_2^3} -\frac{3(1-6\mathcal{T})}{5T_2}  -\frac{3T_2}{5}, \non \\ b_2^{(8)} &=& \frac{8(1-\mathcal{T})}{5T_2} +\frac{8T_2}{5}.\eea
Here $b_1^{(8)}$ and $b_2^{(8)}$ satisfy the Poisson equations
\bea \label{source}\Delta_T b_1^{(8)} &=& 12 b_1^{(8)} +\frac{36 T_2}{5} \delta(T_1), \non \\ \Delta_T b_2^{(8)} &=& 2 b_2^{(8)} -\frac{16 T_2}{5} \delta(T_1),\eea
where we have used
\be \frac{\p \vert x\vert}{\p x} = {\rm sign}(x), \quad \frac{\p {\rm sign}(x)}{\p x} = 2\delta(x).\ee
Thus
\be \label{break1} \mathcal{I}^{D^8\mathcal{R}^4}_{finite} = \mathcal{I}^{D^8\mathcal{R}^4}_1 +\mathcal{I}^{D^8\mathcal{R}^4}_2,\ee
where $\mathcal{I}^{D^8\mathcal{R}^4}_i (i=1,2)$ satisfy the Poisson equations
\bea \label{Pe}\Delta_\Omega \mathcal{I}^{D^8\mathcal{R}^4}_1 &=& 12 \mathcal{I}^{D^8\mathcal{R}^4}_1 +\frac{80}{81\pi^9} E_{5/2}^2, \non \\ \Delta_\Omega \mathcal{I}^{D^8\mathcal{R}^4}_2 &=& 2 \mathcal{I}^{D^8\mathcal{R}^4}_2 - \frac{9}{20 \pi^9} E^2_{5/2}, \eea
where the source terms in \C{source} contributions are obtained using
\bea &&\sum_{{}^{(m_1,m_2)\neq(0,0)}_{(n_1,n_2)\neq (0,0)}} \int_0^\infty d y y^4 \int_1^\infty \frac{dT_2}{T_2}  e^{-\pi^2 y \Big(\vert m_1 + m_2 \Omega\vert^2 T_2 + \vert n_1 +n_2 \Omega \vert^2 T_2^{-1}\Big)/\Omega_2} \non \\ &&= \frac{1}{4}\sum_{{}^{(m_1,m_2)\neq(0,0)}_{(n_1,n_2)\neq (0,0)}} \int_0^\infty dx x^{3/2} e^{-\pi^2 \vert m_1 +m_2\Omega \vert^2 x/\Omega_2}\int_0^\infty dz z^{3/2} e^{-\pi^2 \vert n_1 +n_2\Omega \vert^2 z/\Omega_2}\non \\ &&= \frac{9}{64 \pi^9} E_{5/2}^2,\eea 
where we have substituted 
\be x = yT_2, \quad z= y/T_2\ee
in the intermediate step.

Thus upto an overall numerical factor, these contributions lead to a term in the 9 dimensional effective action
\be \label{EF1}l_{11}^7 \int d^9 x \sqrt{-G^{(9)}} \mathcal{V}_2^{-11/2} E_{3/2} (\Omega,\bar\Omega)\mathcal{I}^{D^8\mathcal{R}^4}(\Omega,\bar\Omega) D^8\mathcal{R}^4.\ee
Based on our analysis of calculations in two loop supergravity in appendix E for special values of the moduli, we expect this contribution to be a source term for the exact amplitude.

Consider the perturbative contributions to $\mathcal{I}^{D^8\mathcal{R}^4}_1$ and $\mathcal{I}^{D^8\mathcal{R}^4}_2$ in \C{Pe}. Ignoring the various non--perturbative contributions and using
\bea \label{E5/2}E_{5/2} = 2\zeta(5) \Omega_2^{5/2} +\frac{8}{3} \zeta(4) \Omega_2^{-3/2}+\ldots,\eea
we get that
\bea \label{A1}\mathcal{I}^{D^8\mathcal{R}^4}_1 &=& \frac{a_0}{\pi^9} \Omega_2^4 +\frac{a_1}{\pi^9} \Omega_2^{-3}-\frac{80}{81\pi^9}\Big(\frac{8}{9} \zeta(4) \zeta(5)\Omega_2 -\frac{1}{2} \zeta(5)^2 \Omega_2^5 +\frac{64}{63} \zeta(4)^2 \Omega_2^{-3} {\rm ln}\Omega_2 \Big) , \non \\ \mathcal{I}^{D^8\mathcal{R}^4}_2 &=& \frac{b_0}{\pi^9} \Omega_2^2 + \frac{b_1}{\pi^9} \Omega_2^{-1} -\frac{9}{20\pi^9} \Big(\frac{32}{45} \zeta(4)^2\Omega_2^{-3} - \frac{16}{3} \zeta(4)\zeta(5)\Omega_2 +\frac{2}{9} \zeta(5)^2 \Omega_2^5\Big)  \eea
where $a_0, a_1, b_0, b_1$ are arbitrary constants. 

We also make use of the expression
\be \label{E3/2}E_{3/2} = 2\zeta(3) \Omega_2^{3/2} +4 \zeta(2) \Omega_2^{-1/2}+\ldots\ee
in evaluating the perturbative contributions in \C{EF1}. 

The primary aim of obtaining these perturbative contributions is to see how the structure of supergravity ties in with the structure of perturbative string theory. This is because the terms in the effective action that have been obtained by compactifying M theory on $T^2$ also yield terms in the effective action of the IIA and IIB theories compactified on $S^1$. The relevant details of the map are given in appendix C. Thus our analysis leads to several perturbative contributions to the $D^8\mathcal{R}^4$ interaction in the  type II effective action. However \C{EF1} leads to several terms in the type II effective action that cannot be interpreted perturbatively because of their dilaton dependence. Hence the coefficient of these terms must vanish in the total amplitude, imposing strong constraints on the supergravity amplitude\footnote{This is also the case where the complete amplitude can be evaluated exactly. For example, the ultraviolet divergent parts of the three loop amplitude also have such contributions after regularizing~\cite{Basu:2014uba}. Note that this does not happen at two loops.}. From \C{EF1}, we now write down terms in the IIB effective action that are consistent with perturbative string theory. We ignore terms that do not have a perturbative interpretation. Thus \C{EF1} gives us (ignoring an overall factor of $\pi^{-9}$)\footnote{We have ignored a term involving 
\be r_B^5 \zeta(3) \zeta(4) \zeta(5) e^{-2\phi^B}\ee
as this is inconsistent with the genus zero amplitude, which is of the form
\be r_B \zeta(7) e^{-2\phi^B}.\ee}
\bea && l_s^7 \int d^9 x \sqrt{-g^{B}} r_B^5 \Big[ 2\zeta(3) b_1 + \frac{22192}{3645} \zeta(2) \zeta(4)\zeta(5) \non \\ && +2\zeta(3) \Big(a_1 - \frac{8}{25} \zeta(4)^2\Big) e^{2\phi^B} + 4b_1 \zeta(2) e^{2\phi^B}  -\frac{10240}{5103} \zeta(3) \zeta(4)^2 e^{2\phi^B} {\rm ln}(e^{-\phi^B}) \non \\ && +4\zeta(2)\Big(a_1 - \frac{8}{25} \zeta(4)^2\Big) e^{4\phi^B} -\frac{20480}{5103} \zeta(2) \zeta(4)^2 e^{4\phi^B} {\rm ln}(e^{-\phi^B})\Big] D^8\mathcal{R}^4.\eea
This leads to several possible contributions at genera 1, 2 and 3 in string theory. Of course whether these contributions are there or not in the various string amplitudes depend on the complete answer for the $D^8\mathcal{R}^4$ interaction. Our analysis does not further constrain these amplitudes. Note that there are some terms of the form ${\rm ln} (e^{-\phi^B})$. This refers to the presence of non--local infra red divergent terms in the effective action which arise from the propagation of massless modes in the loops in the various string amplitudes. These lead to terms of the schematic form $s^4{\rm ln} (-\alpha' \mu_4 s) \mathcal{R}^4$ in the string amplitudes, where $\mu_4$ is uniquely determined in string theory, which shows up as a moduli dependent scale in supergravity\footnote{This is also a feature of two loop maximal supergravity~\cite{Green:2008bf}. Of course, the scale is dilaton independent in the string frame in string theory.}. As we see, this scale dependence also shows up in the analysis of the local terms.   

Thus we see that $a_0$ and $b_0$ yield contributions that are inconsistent with the structure of string perturbation theory, and hence they must vanish. The coefficients $a_1$ and $b_1$ are calculated in appendix D. Thus from the explicit expression for the amplitude we get that for $A=0+ A=1$ in \C{mostimp} (hence multiplying by a factor of 2)
\bea \label{form1}
\mathcal{A}_4^{(3)} &=& (2\pi^8 l_{11}^{15} r_B) \Sigma_4 \mathcal{R}^4 \frac{l_s^8 r_B^5}{45\cdot 729}\Big[\frac{243}{56} \zeta(2)\zeta(3)^2 +\frac{1387}{2^5\cdot 5} \zeta(2) \zeta(5) \non \\&& + \frac{5}{7} \zeta(3)\zeta(4) e^{2\phi^B}\Big(\frac{893437}{588000} +{\rm ln} (\mu e^{4\phi^B}) \Big)  \non \\ && -\frac{10}{7}\zeta(2)\zeta(4) e^{4\phi^B}\Big(\frac{127163}{84000} -{\rm ln} (\mu e^{4\phi^B}) \Big) \Big],\eea 
where
\be \label{defmu}{\rm ln} \mu = {\rm ln}(16e^{-\gamma}) +\frac{8\zeta'(4)}{\zeta(4)} - \frac{4\zeta'(8)}{\zeta(8)}.\ee

The overall factor of $2\pi^8 l_{11}^{15} r_B$ provides the correct normalization to the amplitude and is common to all loop amplitudes. Note that the scale of the logarithmic terms in the amplitude involve ratios of the form $\zeta'(s)/\zeta(s)$ for integral $s$. This is exactly similar to the structure of the scale of the logarithmic terms obtained in the low energy expansion of the genus one four graviton amplitude~\cite{Green:2008uj}\footnote{This feature also persists for the $D^{10} \mathcal{R}^4$ amplitude, as we shall see later.}. It is plausible that this feature survives at all genera for the string amplitude. In \C{form1}, the genus one contribution of the form $\zeta(2)\zeta(3)^2$ agrees with the structure obtained in~\cite{Green:2008uj}. 

For the $D^8\mathcal{R}^4$ amplitude (as well as the $D^{10} \mathcal{R}^4$ amplitude as we shall see later) the terms of the form ${\rm ln}(e^{-\phi^B})$ in the type IIB theory yield terms of the form ${\rm ln}(r_A e^{-\phi^A})$ in the type IIA theory on using \C{relate}. Hence these lead to contributions logarithmic in the radius of the compactification in 9 dimensions. These contributions indeed arise in string theory (see~\cite{Green:2008uj} for examples at genus one). In fact perturbative equality of the four graviton amplitude upto genus three implies that such contributions must also be there in the type IIB theory if they are there in the type IIA theory.

We can also perform the exact analysis when $T_1 =0$ in \C{mostimp}. This corresponds to the left boundary in the $T$ plane in figure 6. The details of the analysis are given in appendix B. 

\section{The structure of the three loop $D^{10}\mathcal{R}^4$ amplitude}

We now consider the structure of the three loop $D^{10} \mathcal{R}^4$ amplitude in detail. Since the analysis is very similar to the one for the $D^8\mathcal{R}^4$ amplitude, we only give the results and skip several details. Consider the expression
\be \int d\mu f_2 (L,T_1,T_2,A) \hat{F}_L\ee 
where $f_2$ is given in \C{mostimp2} integrated over the domain \C{constraint}. Starting from the relevant region in the $T$ plane in figure 6, we now map it to patches of $\mathcal{F}_T$ as before.   

The contributions from $h_1 \oplus f_2$ add to give
\be \int_{\mathcal{F}_T} d\mu f_2^{(1)} (L,\vert T_1\vert,T_2,A) \hat{F}_L (L,T,A) \ee
along with the constraints \C{contone},
where
\bea f_2^{(1)} (L,T_1,T_2,A)&=& \Big[ -1 + \frac{\vert T\vert^2}{2T_2} \Big( \frac{1}{L^3} + \frac{\vert T\vert^2}{T_2}\Big) + \frac{4\vert T \vert^2}{T_2} \Big( \frac{1-T_1}{T_2} +\frac{A(A-1)}{L^3}\Big) \non \\ &&-\frac{\vert T\vert^4}{T_2^2}\Big( \frac{1-T_1}{T_2} +\frac{A(A-1)}{L^3}\Big)^2 \Big] \Big[ (1-T_1)\frac{L}{T_2} + \frac{A(A-1)}{L^2 }\Big] \non \\ && \times \frac{1}{L T_2}\Big[ AT_1 + (1-A)(\vert T\vert^2 - T_1)\Big].\eea
The contributions from $f_1 \oplus g_1$ add to give
\be \int_{\mathcal{F}_T} d\mu f_2^{(2)} (L,\vert T_1\vert,T_2,A) \hat{F}_L (L,1/(1-T),A) \ee
along with the constraints \C{conttwo},
where
\bea f_2^{(2)} (L,T_1,T_2,A)&=& \Big[ -1 + \frac{1}{2T_2} \Big( \frac{1}{L^3} + \frac{1}{T_2}\Big) + \frac{4}{T_2} \Big( \frac{\vert T\vert^2-T_1}{T_2} +\frac{A(A-1)}{L^3}\Big) \non \\ &&-\frac{1}{T_2^2}\Big( \frac{\vert T\vert^2-T_1}{T_2} +\frac{A(A-1)}{L^3}\Big)^2 \Big] \Big[ (\vert T\vert^2-T_1)\frac{L}{T_2} + \frac{A(A-1)}{L^2 }\Big] \non \\ && \times \frac{1}{L T_2}\Big[ A(1-T_1) + (1-A)T_1\Big].\eea
Finally the contributions from $g_2 \oplus h_2$ add to give
\be \int_{\mathcal{F}_T} d\mu f_2^{(3)} (L,\vert T_1\vert,T_2,A) \hat{\mathcal{F}}_L (L,\vert T_1 \vert, T_2,A) \ee
along with the constraints \C{contthree},
where
\bea f_2^{(3)} (L,T_1,T_2,A)&=& \Big[ -1 + \frac{\vert T-1\vert^2}{2T_2} \Big( \frac{1}{L^3} + \frac{\vert T-1\vert^2}{T_2}\Big) + \frac{4\vert T-1\vert^2}{T_2} \Big( \frac{T_1}{T_2} +\frac{A(A-1)}{L^3}\Big) \non \\ &&-\frac{\vert T-1\vert^4}{T_2^2}\Big( \frac{T_1}{T_2} +\frac{A(A-1)}{L^3}\Big)^2 \Big] \Big[ \frac{T_1L}{T_2} + \frac{A(A-1)}{L^2 }\Big] \non \\ && \times \frac{1}{L T_2}\Big[ A(\vert T \vert^2-T_1) + (1-A)(1-T_1)\Big].\eea
Again using the symmetry of $A \rightarrow 1-A$, we get that the total contribution is equal to
\be D_1 + D_2 + D_3,\ee
where

{\bf{(i)}}  \be D_1 = \int_{\mathcal{F}_T} d\mu G_1 (L,\vert T_1\vert,T_2,A)\hat{F}_L (L,T,A)\ee
where
 \bea G_1 (L,\vert T_1 \vert ,T_2,A)&=& \frac{\vert T\vert^2}{2T_2}\Big[ -1 + \frac{\vert T\vert^2}{2T_2} \Big( \frac{1}{L^3} + \frac{\vert T\vert^2}{T_2}\Big) + \frac{4\vert T \vert^2}{T_2} \Big( \frac{1-\vert T_1 \vert}{T_2} +\frac{A(A-1)}{L^3}\Big) \non \\ &&-\frac{\vert T\vert^4}{T_2^2}\Big( \frac{1-\vert T_1 \vert}{T_2} +\frac{A(A-1)}{L^3}\Big)^2 \Big] \Big[ \frac{(1-\vert T_1 \vert)}{T_2} + \frac{A(A-1)}{L^3 }\Big] \eea
along with the constraints \C{contone},

{\bf{(ii)}} \be D_2 = \int_{\mathcal{F}_T} d\mu G_2 (L,\vert T_1\vert,T_2,A)\hat{F}_L (L,1/(1-T),A)\ee
where
\bea G_2 (L,\vert T_1 \vert,T_2,A)&=& \frac{1}{2T_2}\Big[ -1 + \frac{1}{2T_2} \Big( \frac{1}{L^3} + \frac{1}{T_2}\Big) + \frac{4}{T_2} \Big( \frac{\vert T\vert^2-\vert T_1 \vert}{T_2} +\frac{A(A-1)}{L^3}\Big) \non \\ &&-\frac{1}{T_2^2}\Big( \frac{\vert T\vert^2-\vert T_1 \vert}{T_2} +\frac{A(A-1)}{L^3}\Big)^2 \Big] \Big[ \frac{(\vert T\vert^2-\vert T_1 \vert)}{T_2} + \frac{A(A-1)}{L^3 }\Big] \non \\ \eea
along with the constraints \C{conttwo}, and

{\bf{(iii)}} \be D_3 = \int_{\mathcal{F}_T} d\mu G_3 (L,\vert T_1\vert,T_2,A)\hat{\mathcal{F}}_L (L,\vert T_1 \vert, T_2,A)\ee
where
\bea G_3 (L,\vert T_1 \vert,T_2,A)&=& \Big[ -1 + \frac{\vert T\vert^2 - 2\vert T_1 \vert +1}{2T_2} \Big( \frac{1}{L^3} + \frac{\vert T\vert^2 - 2\vert T_1 \vert +1}{T_2}\Big) \non \\ && + \frac{4(\vert T\vert^2 - 2\vert T_1 \vert +1)}{T_2} \Big( \frac{\vert T_1 \vert }{T_2} +\frac{A(A-1)}{L^3}\Big) \non \\ &&-\frac{(\vert T\vert^2 - 2\vert T_1 \vert +1)^2}{T_2^2}\Big( \frac{\vert T_1 \vert}{T_2} +\frac{A(A-1)}{L^3}\Big)^2 \Big] \Big[ \frac{\vert T_1 \vert}{T_2} + \frac{A(A-1)}{L^3 }\Big] \non \\ && \times \frac{( \vert T\vert^2 - 2\vert T_1 \vert +1)}{2T_2}.\eea
along with the constraints \C{contthree}.

Thus the unrenormalized $D^{10} \mathcal{R}^4$ interaction is given by
\be \label{veryimp}I^{D^{10} \mathcal{R}^4} = \frac{\pi^{21/2} \Sigma_5}{768 l_{11}^{11}} \int_0^\infty dV_3 V_3^{8/3} (D_1 + D_2 + D_3).\ee

\subsection{$A=0 +A=1$ in \C{mostimp2}}

As before, the total contribution to $A=0$ and $A=1$ together in \C{mostimp2} is given by twice the contribution from \C{veryimp} for $A=0$.  To evaluate it, from \C{veryimp} we see that
\bea \label{10fin}
I^{D^{10} \mathcal{R}^4} = \frac{\pi^{21/2} \Sigma_5}{768 l_{11}^{11}} \int_0^\infty dV_3 V_3^{8/3} \int_0^\infty \frac{dL}{L^2} \int_{\mathcal{F}_T} \frac{dT_1 dT_2}{T_2^2} F^{D^{10} \mathcal{R}^4} (T,\bar{T},L)  \hat{F}_L (L,T),\eea
where
\be F^{D^{10} \mathcal{R}^4} (T,\bar{T},L) = \frac{F^{D^8 \mathcal{R}^4}(T,\bar{T})}{4L^3} + F^{new} (T,\bar{T})\ee
on using \C{FD8R4}, where
\bea && F^{new} (T,\bar{T}) = -1 + \frac{\vert T \vert^4 (1-\vert T_1 \vert)}{2 T_2^2} \Big[ \frac{\vert T \vert^2}{2 T_2^2} +\frac{4(1-\vert T_1 \vert)}{T_2^2} - \frac{\vert T \vert^2 (1- \vert T_1\vert)^2}{T_2^4}\Big]  \non \\ && +\frac{\vert T\vert^2 - \vert T_1 \vert}{2 T_2^2} \Big[ \frac{1}{2 T_2^2} +\frac{4(\vert T\vert^2 - \vert T_1\vert)}{T_2^2} - \frac{ (\vert T\vert^2- \vert T_1\vert)^2}{T_2^4}\Big] \non \\ &&+\frac{\vert T_1\vert (\vert T\vert^2 - 2\vert T_1\vert+1)^2}{2T_2^2} \Big[ \frac{\vert T\vert^2 - 2\vert T_1\vert+1}{2 T_2^2} +\frac{4\vert T_1 \vert}{T_2^2} - \frac{\vert T_1\vert^2 (\vert T\vert^2 - 2\vert T_1\vert+1)}{T_2^4}\Big]\non \\ \eea
which is independent of $L$.

We now evaluate the finite part of \C{10fin}, given that
\be I^{D^{10} \mathcal{R}^4} =b_0 \Lambda^{11} +\ldots + (l_{11}^2 \mathcal{V}_2)^{-11/2} g(\Omega,\bar\Omega),\ee
where $b_0$ is moduli independent. This gives us that
\bea \label{def2}
 I^{D^{10}\mathcal{R}^4} &=& \frac{\pi^{10}E_{1/2} (\Omega,\bar\Omega)\Sigma_5 }{6144 (l_{11}^2 \mathcal{V}_2)^{11/2}} \mathcal{I}^{D^8\mathcal{R}^4} (\Omega,\bar\Omega) +  \frac{\pi^8 E_{3/2} (\Omega,\bar\Omega)\Sigma_5 }{3072 (l_{11}^2 \mathcal{V}_2)^{11/2}} \mathcal{I}^{new} (\Omega,\bar\Omega) \eea
where $\mathcal{I}^{D^8\mathcal{R}^4}$ is defined in \C{A01} and evaluated in \C{break1}. Also
\bea \mathcal{I}^{new} = \int_0^\infty d y y^3 \int_{\mathcal{F}_T} \frac{dT_1 dT_2}{T_2^2} F^{new} (T,\bar{T}) \sum_{\hat{m}^I,\hat{n}^I} e^{-\pi^2 y \hat{G}_{IJ} (\hat{m} +\hat{n}T)^I(\hat{m}+\hat{n}\bar{T})^J/T_2}    . \eea
Thus
\bea \Delta_\Omega \mathcal{I}^{new}_{finite} = \int_0^\infty d y y^3 \int_{\mathcal{F}_T} \frac{dT_1 dT_2}{T_2^2} \Delta_T F^{new} (T,\bar{T}) \sum_{{}^{(\hat{m}^1,\hat{m}^2)\neq(0,0)}_{(\hat{n}^1,\hat{n}^2)\neq (0,0)}} e^{-\pi^2 y \hat{G}_{IJ} (\hat{m} +\hat{n}T)^I(\hat{m}+\hat{n}\bar{T})^J/T_2}    .\non \\ \eea
Hence we need to analyze the structure of $\Delta_T F^{new}$. Using \C{calT}, we get that
\be F^{new} (T,\bar{T}) = \frac{3\mathcal{T}^4}{2T_2^6} +\frac{\mathcal{T}^2(5-17\mathcal{T})}{4T_2^4} +\frac{1-7\mathcal{T}+17\mathcal{T}^2}{4T_2^2} +\frac{8-7\mathcal{T}}{4} +\frac{T_2^2}{4}.\ee
This leads to
\be F^{new} (T,\bar{T})= b_1^{(10)} (T,\bar{T})  + b_2^{(10)} (T,\bar{T}) +b_3^{(10)} (T,\bar{T})+b_4^{(10)} (T,\bar{T}),\ee
where $b_i^{(10)}$ ($i=1,2,3,4$) are given by 
\bea 
b_1^{(10)} &=& \frac{3\mathcal{T}^4}{2T_2^6} +\frac{3\mathcal{T}^2(3 -14\mathcal{T})}{11 T_2^4} +\frac{1-20\mathcal{T} +70\mathcal{T}^2}{22T_2^2} +\frac{5(3-14\mathcal{T})}{77} +\frac{T_2^2}{22},\non \\ b_2^{(10)} &=& -\frac{19\mathcal{T}^2(\mathcal{T}-1)}{44 T_2^4} +\frac{19(1-9\mathcal{T}+15\mathcal{T}^2)}{308T_2^2} +\frac{57(4-15\mathcal{T})}{1540} +\frac{19T_2^2}{308}, \non \\
b_3^{(10)} &=& \frac{(\mathcal{T}-1)^2}{7T_2^2} +\frac{1-2\mathcal{T}}{7} +\frac{T_2^2}{7}, \non \\ b_4^{(10)} &=& \frac{53}{35}.
\eea
Now $b_i^{(10)}$ satisfy the Poisson equations
\bea \Delta_T b_1^{(10)} &=& 42 b_1^{(10)} - \frac{20}{11} T_2 (T_2 + T_2^{-1}) \delta(T_1), \non \\ \Delta_T b_2^{(10)} &=& 20 b_2^{(10)} - \frac{171}{154} T_2 (T_2 + T_2^{-1}) \delta(T_1), \non \\ \Delta_T b_3^{(10)} &=& 6 b_3^{(10)} - \frac{4}{7} T_2 (T_2 + T_2^{-1}) \delta(T_1), \non \\ \Delta_T b_4^{(10)} &=& 0.\eea
Thus from the contributions $b_i^{(10)} (i=1,2,3)$, we get that
\be \mathcal{I}^{new}_{finite} = \mathcal{I}_1^{new} + \mathcal{I}_2^{new} +\mathcal{I}_3^{new},\ee
where $\mathcal{I}_i^{new}$ satisfy the Poisson equations
\bea \label{Pe2}\Delta_\Omega \mathcal{I}_1^{new} &=& 42 \mathcal{I}_1^{new} - \frac{15}{44\pi^7}E_{3/2} E_{5/2}, \non \\ \Delta_\Omega \mathcal{I}_2^{new} &=&20\mathcal{I}_2^{new} -\frac{513}{2464\pi^7} E_{3/2} E_{5/2}, \non \\ \Delta_\Omega \mathcal{I}_3^{new} &=&6 \mathcal{I}_3^{new} - \frac{3}{28\pi^7}E_{3/2} E_{5/2}.\eea
The remaining contribution to $\mathcal{I}^{new}$ comes from the zero mode involving $b_4^{(10)}$, which we evaluate directly. This contribution is given by
\bea \mathcal{I}^{new}_{0-mode} = \frac{53}{35}\int_0^\infty d y y^3 \int_{\mathcal{F}_T} \frac{dT_1 dT_2}{T_2^2}  \sum_{\hat{m}^I,\hat{n}^I} e^{-\pi^2 y \hat{G}_{IJ} (\hat{m} +\hat{n}T)^I(\hat{m}+\hat{n}\bar{T})^J/T_2}. \eea    
The finite part is given by the non--degenerate orbits of $SL(2,\mathbb{Z})$ and is equal to
\bea \mathcal{I}^{new}_{0-mode}= \frac{53\zeta(3)\zeta(4)}{35\pi^7}\eea
on using \C{int}. 

Thus upto an overall numerical factor, these contributions lead to a term in the 9 dimensional effective action
\be l_{11}^9 \int d^9 x \sqrt{-G^{(9)}} \mathcal{V}_2^{-9/2} \Big[ \frac{\pi^2}{2} E_{1/2} (\Omega,\bar\Omega)\mathcal{I}^{D^8\mathcal{R}^4}(\Omega,\bar\Omega) + E_{3/2} (\Omega,\bar\Omega) \mathcal{I}^{new}(\Omega,\bar\Omega) \Big] D^{10}\mathcal{R}^4.\ee
Again, based on our analysis of calculations in two loop supergravity in appendix E for special values of the moduli, we expect this contribution to yield source terms for the exact amplitude.

Let us now consider the perturbative contributions to $\mathcal{I}^{new}_1$, $\mathcal{I}^{new}_2$ and $\mathcal{I}^{new}_3$ in \C{Pe2}. Using the perturbative parts of $E_{3/2}$ and $E_{5/2}$ given by \C{E3/2} and \C{E5/2} respectively, we get that
\bea \label{c1} \mathcal{I}^{new}_1 &=& \frac{c_0}{\pi^7} \Omega_2^7 +\frac{c_1}{\pi^7} \Omega_2^{-6} + \frac{15}{44\pi^7} \Big(\frac{8}{27}\zeta(2)  \zeta(4) \Omega_2^{-2} +\frac{8}{63} \zeta(3)\zeta(4)+ \frac{1}{5} \zeta(2) \zeta(5) \Omega_2^2 \non \\ &&+ \frac{2}{15} \zeta(3)\zeta(5)\Omega_2^4 \Big), \non \\ \mathcal{I}^{new}_2 &=& \frac{d_0}{\pi^7} \Omega_2^5 +\frac{d_1}{\pi^7} \Omega_2^{-4}+\frac{513}{2464\pi^7} \Big(\frac{16}{21}\zeta(2)  \zeta(4) \Omega_2^{-2} +\frac{4}{15} \zeta(3)\zeta(4) + \frac{4}{9} \zeta(2) \zeta(5) \Omega_2^2  \non \\ &&+ \frac{1}{2} \zeta(3)\zeta(5)\Omega_2^4\Big), \non \\ \mathcal{I}^{new}_3 &=& \frac{e_0}{\pi^7} \Omega_2^3 +\frac{e_1}{\pi^7} \Omega_2^{-2} +\frac{3}{28\pi^7} \Big(\frac{8}{9} \zeta(3)\zeta(4) +2\zeta(2)\zeta(5)\Omega_2^2 -\frac{2}{3} \zeta(3)\zeta(5) \Omega_2^4 \non \\ &&+\frac{32}{15} \zeta(2) \zeta(4) \Omega_2^{-2} {\rm ln}\Omega_2 \Big), \eea
where $c_0, c_1, d_0, d_1, e_0, e_1$ are arbitrary constants.

These lead to various terms in the type IIB theory that are consistent with the structure of perturbative string theory. Using the perturbative part of $E_{1/2}$~\cite{Green:2008bf}
\be E_{1/2} = 2\sqrt{\Omega_2} {\rm ln} \Big( \frac{\Omega_2}{4\pi e^{-\gamma}} \Big)  \ee
and $E_{3/2}$ in \C{E3/2} they are (ignoring an overall factor of $\pi^{-7}$)\footnote{We have ignored terms involving
\be r_B^3 \zeta(2)\zeta(3)\Big( 5\zeta(4) +12\zeta(5) \Big) e^{-2\phi^B}\ee 
as the genus zero amplitude is of the form
\be r_B \zeta(3) \zeta(5)e^{-2\phi^B}.\ee}
\bea &&l_s^9 \int d^9 x \sqrt{-g^B} r_B^3 {\rm ln} (e^{-\phi^B}/4\pi e^{-\gamma})\Big[ \frac{5548}{3645} \zeta(4) \zeta(5) + b_1 e^{2\phi^B} + \Big(a_1 -\frac{8}{25} \zeta(4)^2\Big) e^{4\phi^B} \non \\ && - \frac{5120}{5103}\zeta(4)^2 e^{4\phi^B} {\rm ln}(e^{-\phi^B})\Big] D^{10} \mathcal{R}^4\non \\ &&+l_s^9 \int d^9 x \sqrt{-g^B} r_B^3 \Big[ \frac{41}{12} \zeta(3)^2 \zeta(4) + \frac{3}{2} \zeta(2)^2 \zeta(5) + 2 \Big(e_1 +\frac{229}{882} \zeta(2) \zeta(4)\Big) \zeta(3) e^{2\phi^B}\non \\ && + \frac{41}{6} \zeta(2) \zeta(3)\zeta(4) e^{2\phi^B} + \frac{16}{35} \zeta(2) \zeta(3) \zeta(4) e^{2\phi^B} {\rm ln} (e^{-\phi^B}) + 2 d_1 \zeta(3) e^{4\phi^B} \non \\ &&+ 4 \Big( e_1 +\frac{229}{882} \zeta(2) \zeta(4) \Big)\zeta(2) e^{4\phi^B} +\frac{32}{35} \zeta(2)^2 \zeta(4) e^{4\phi^B} {\rm ln}(e^{-\phi^B}) + 2 c_1 \zeta(3) e^{6\phi^B} \non \\ &&+ 4 d_1 \zeta(2) e^{6\phi^B} + 4 c_1 \zeta(2) e^{8\phi^B}\Big] D^{10} \mathcal{R}^4. \eea
This leads to several possible contributions at genera 1, 2, 3, 4 and 5 in string theory. These include logarithmic terms like the $D^8\mathcal{R}^4$ interaction. Now $c_0, d_0$ and $e_0$ yield contributions that are inconsistent with the structure of string perturbation theory, and hence they must vanish. The coefficients $c_1, d_1$ and $e_1$ are calculated in appendix D.   
Thus for $A=0 +A=1$ from \C{mostimp2} we get that (hence multiplying by a factor of 2)
\bea \label{form2}
&&\mathcal{A}_4^{(3)} = (2\pi^8 l_{11}^{15} r_B) \Sigma_5 \mathcal{R}^4 \frac{l_s^{10} r_B^3}{63 \cdot 96} {\rm ln} (e^{-\phi^B}/4\pi e^{-\gamma})\Big[ \frac{7\cdot 1387}{ 60\cdot 405} \zeta(2) \zeta(5) + \zeta(3)\zeta(4) e^{2\phi^B} \non \\ &&- \frac{80}{15\cdot 81} \zeta(2)\zeta(4)\Big(\frac{127163}{84000} -{\rm ln} (\mu e^{4\phi^B}) \Big) e^{4\phi^B} \Big] \non \\ &&+(2\pi^8 l_{11}^{15} r_B) \Sigma_5 \mathcal{R}^4 \frac{l_s^{10} r_B^3}{15\cdot 1536} \Big[ \frac{41}{12} \zeta(2) \zeta(3)^2 +\frac{15}{4} \zeta(2) \zeta(5)  \non \\ &&+\frac{8}{7}\zeta(3)\zeta(4)\Big(\frac{166661}{10080}  +  {\rm ln} (\nu e^{-\phi^B}) \Big)e^{2\phi^B} +\frac{16}{7}\zeta(2) \zeta(4)\Big(\frac{7993}{5040}  +  {\rm ln} (\nu e^{-\phi^B}) \Big)e^{4\phi^B} \non \\ && +\frac{19}{280} \zeta(3) \zeta(6) e^{4\phi^B}  +\frac{8}{21 \cdot 99} \zeta(3)\zeta(8) e^{6\phi^B} + \frac{19}{140} \zeta(8) e^{6\phi^B} + \frac{4}{315} \zeta(10) e^{8\phi^B}\Big] ,\non \\ \eea 
where
\be {\rm ln} \nu = {\rm ln}(e^{\gamma}/2) -\frac{\zeta'(2)}{\zeta(2)} - \frac{\zeta'(4)}{\zeta(4)} +\frac{\zeta'(6)}{\zeta(6)},\ee
and ${\rm ln}  \mu$ is defined in \C{defmu}. In \C{form2}, the genus one contribution of the form $\zeta(2)\zeta(3)^2$ agrees with the structure obtained in~\cite{Green:2008uj}.   

\section{The four loop ladder diagram contribution to the four graviton amplitude}

At four loops the four graviton amplitude is given by~\cite{Bern:2009kd}
\bea \mathcal{A}_4^{(4)} = \frac{(4\pi^2)^4 \kappa_{11}^{10}}{(2\pi)^{44}}\sum_{S_4} \sum_{i=1}^{50} c_i I^{(i)}  \mathcal{K},\eea
where $S_4$ is the set of 24 permutations of the massless external legs $\{1,2,3,4\}$, and $c_i$ are constants. There are 50 diagrams that contribute at four loops, each of which is constructed using the skeleton diagrams in figure 4. Note that the ladder skeleton is given by $b$ in figure 4. We shall consider the leading term in the low momentum expansion of the four loop amplitude that arises from the ladder skeleton. 

The leading contribution in the low momentum expansion is the $O(D^8\mathcal{R}^4)$ amplitude. Among all the diagrams that follow from the ladder diagram, only the diagram 3 (in the conventions of~\cite{Bern:2009kd}) contributes at this order. This non--planar diagram is given by figure 8, where $l_i$ denote the momenta along the corresponding link in the loop diagram.

\begin{figure}[ht]
\begin{center}
\[
\mbox{\begin{picture}(200,110)(0,0)
\includegraphics[scale=.5]{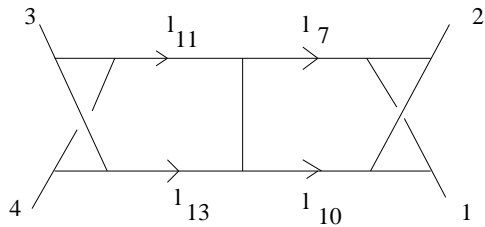}
\end{picture}}
\]
\caption{The four loop ladder diagram contribution}
\end{center}
\end{figure}

For this diagram,   
\be c_3 =\frac{1}{16},\ee
while $I^{(3)}$ involves integrating over the loop momenta with massless $\varphi^3$ field theory propagators along with the numerator
\be N^{(3)} = -S^4 \Big[ l_7^2 (l_{11}^2 - 2 l_{13}^2) + l_{10}^2 (l_{13}^2 - 2 l_{11}^2)\Big]\ee
in the integrand.
Since $N^{(3)}$ already has a factor of $S^4$, to extract the $O(D^8\mathcal{R}^4)$ term, we simply evaluate the loop integrals at zero external momenta. Thus in 11 uncompactified dimensions
\be I^{(3)} = 2S^4 \int \frac{d^{11}p_1 d^{11} p_2 d^{11} p_3 d^{11} p_4}{p_1^4 p_2^4 p_3^4 p_4^4 (p_1 + p_2)^2 (p_3 + p_4)^2 (p_1 + p_2 + p_3 + p_4)^2}.\ee 
We now directly write down the answer we get on compactifying on $T^2$ as the analysis is very similar to the earlier ones. We get that
\be I^{(3)} = \frac{2 S^4}{(4\pi^2 l_{11}^2 \mathcal{V}_2)^4} \int_0^\infty d\Xi \s\lambda\rho\mu F_L (\s,\lambda,\rho,\mu,\nu,\theta,\epsilon)\Delta (\s,\lambda,\rho,\mu,\nu,\theta,\epsilon),\ee
where the measure involves
\be d\Xi = d\s d\lambda d\rho d\mu d\nu d\theta d\epsilon. \ee 
The lattice sum involving the KK momenta is given by
\be F_L (\s,\lambda,\rho,\mu,\nu,\theta,\epsilon) = \sum_{l_I,m_I,n_I,r_I} e^{-\Big(\s {\bf{l}}^2 +\lambda {\bf{m}}^2 +\rho {\bf{n}}^2 +\mu {\bf{r}}^2 +\nu ({\bf{l+m}})^2 +\theta ({\bf{n+r}})^2 +\epsilon ({\bf{l+m+n+r}})^2 \Big)/l_{11}^2},\ee
while the contribution from the non--compact momenta is given by
\bea &&\Delta (\s,\lambda,\rho,\mu,\nu,\theta,\epsilon) \non \\ &&= \int_0^\infty d^9 p_1 d^9 p_2 d^9 p_3 d^9 p_4 e^{-\Big( \s p_1^2 + \lambda  p_2^2 + \rho p_3^2 + \mu p_4^2 + \nu (p_1 + p_2)^2 +\theta (p_3 + p_4)^2 +\epsilon(p_1 + p_2 + p_3 + p_4)^2\Big)}.\non \\ \eea
We write the lattice factor in a compact way as 
\be F_L (\s,\lambda,\rho,\mu,\nu,\theta,\epsilon) = \sum_{k_{\alpha I}} e^{-G^{IJ} G^{\alpha\beta} k_{\alpha I} k_{\beta J}/l_{11}^2}\ee
where the KK integers $k_{\alpha I}$ are defined by $k_{\alpha I} = \{l_I, m_I,n_I,r_I\}$ for $\alpha = 1,2,3,4$. Thus we have that
\be \label{inverse2}G^{\alpha\beta} = \begin{pmatrix}
\s +\nu +\epsilon &\nu +\epsilon & \epsilon &\epsilon  \\
 \nu +\epsilon & \lambda+\nu+\epsilon & \epsilon &\epsilon \\
\epsilon &\epsilon & \rho+\theta +\epsilon& \theta +\epsilon\\
\epsilon &\epsilon & \theta+\epsilon & \mu +\theta+\epsilon
\end{pmatrix}.\ee  
As in the earlier case, we do not have a detailed understanding of the underlying auxiliary geometry, and so we shall perform the calculation for a special point in the moduli space of the auxiliary geometry. Among the various possible values that are dictated by the structure of \C{inverse2}, we consider the special point $\epsilon =0$, where the inverse metric block diagonalizes. In fact, the entire structure reduces to the product of two amplitudes both of which involve two loop diagrams.

At this special point we get that  
\be I^{(3)} = \frac{2 S^4(l_{11}^2 \mathcal{V}_2)}{(4\pi^2 l_{11}^2 \mathcal{V}_2)^4} \Big[ \pi^9\int_0^\infty d\s d\lambda d\rho \s\lambda \frac{F_L (\s,\lambda,\rho)}{\Delta_2^{9/2} (\s,\lambda,\rho)} \Big]^2,\ee
where the lattice factor $F_L (\s,\lambda,\rho)$ and $\Delta_2(\s,\lambda,\rho)$ are the two loop quantities~\cite{Green:1999pu,Green:2005ba} 
\be F_L (\s,\lambda,\rho) = \sum_{m_I,n_I} e^{-G^{IJ}\Big(\s m_I m_J +\lambda n_I n_J+\rho (m+n)_I (m +n)_J \Big)/l_{11}^2}\ee
and 
\be \Delta_2(\s,\lambda,\rho) = \s\lambda +\lambda\rho+\rho\s.\ee
Thus using the $S_3$ symmetry of the underlying skeleton diagram given in figure 2 (or the dual graph which is the equilateral triangle) we get that
\be I^{(3)} = \frac{2 S^4 \pi^{18}(l_{11}^2 \mathcal{V}_2)}{9(4\pi^2 l_{11}^2 \mathcal{V}_2)^4} \Big[ \int_0^\infty d\s d\lambda d\rho  \frac{F_L(\s,\lambda,\rho)}{\Delta_2^{7/2} (\s,\lambda,\rho)} \Big]^2 .\ee
We proceed exactly as in the two loop analysis. We first perform Poisson resummation, and then express the 3 Schwinger parameters in terms of the volume $y$ and complex structure $T$ of an auxiliary $T^2$. This leads to
\be I^{(3)} = \frac{S^4 \pi^{14}}{32(l_{11}^2 \mathcal{V}_2)^{11}} \mathcal{J}^2,\ee
where
\be \mathcal{J} = \int_0^\infty d y y^5 \int_{\mathcal{F}_T} \frac{dT_1 dT_2}{T_2^2} \hat{F}_L,\ee
where the lattice factor is given by
\be \hat{F}_L = \sum_{\hat{m}^I, \hat{n}^I} e^{-\pi^2 y\hat{G}_{IJ} (\hat{m} +\hat{n} T)^I (\hat{m} +\hat{n} \bar{T})^J /T_2}\ee
which involves a sum over the winding modes. Since $\mathcal{J}$ is $SL(2,\mathbb{Z})_T$ invariant, its finite contribution arises from the non--degenerate orbits of $SL(2,\mathbb{Z})_T$ as described in appendix B, and is given by
\be \mathcal{J}_{finite} = \frac{3\zeta(5)\zeta(6)}{\pi^{11}}.\ee
Thus
\be I^{(3)} = \frac{9S^4 \zeta(5)^2 \zeta(6)^2}{32 \pi^8 (l_{11}^2 \mathcal{V}_2)^{11}}.\ee

Thus upto an overall numerical factor, this leads to a term in the 9 dimensional effective action
\be  \zeta(5)^2 \zeta(6)^2 l_{11}^7 \int d^9 x \sqrt{-G^{(9)}} \mathcal{V}_2^{-10} D^8\mathcal{R}^4.\ee
This fails to provide any non--trivial perturbative contribution as it yields a term proportional to $r_B^{11} e^{-\phi^B}$ in the IIB effective action. 

Based on our calculations above and the general discussion that follows from appendix E, we see the general principle of our analysis. For the special choices of parameters for which we could perform the analysis exactly, we get U--duality invariant contributions to the amplitude that should yield source terms in the Poisson equations of the exact answer, and hence upto numerical factors they capture the perturbative and some features of the non--perturbative parts of the amplitude (including transcendentality) that arise in the exact answer. We expect this logic to hold true at all loops. In particular at three loops, we have looked at the non--BPS $D^8\mathcal{R}^4$ and $D^{10} \mathcal{R}^4$ amplitudes for which very little is known about the Poisson equations they satisfy. Interestingly, our analysis leads to equations \C{A01} and \C{def2} for these amplitudes that generalize the two loop calculations, which involve the $SL(2,\mathbb{Z})$ invariant modular forms $\mathcal{I}^{D^8\mathcal{R}^4}$ and $\mathcal{I}^{new}$ respectively. We expect them to reproduce the structure of the exact amplitude (in the sense mentioned in appendix E) that result from these diagrams. It would be interesting to examine if these modular forms arise as the couplings of some interactions in the effective action.    

While the perturbative parts of our calculation yield results at various genera in perturbative string theory, our results also yield new types of non-perturbative contributions that do not arise from the analysis of one and two loop maximal supergravity. In the IIB theory, these are D--(anti)instanton contributions resulting from bound states of three of them. For the $D^8\mathcal{R}^4$ interaction it follows from the structure of \C{A01}. While one such instanton contribution arises from the non--perturbative part of $E_{3/2}$, the contribution from the other two arises from the non--perturbative part of the source terms involving $E_{5/2}^2$ (which has square of the Bessel functions) in the Poisson equations satisfied by $\mathcal{I}_i^{D^8\mathcal{R}^4}$. Similar is the analysis for the $D^{10} \mathcal{R}^4$ interaction. We expect this analysis to generalize at higher loops, and these non--BPS interactions should receive non--perturbative contributions from bound states involving more and more D--(anti)instantons in the type IIB theory as one goes to higher and higher supergravity loops.

\vspace{.5cm}

{\bf{Acknowledgements:}} 

\vspace{.2cm}

I am thankful to Dileep Jatkar for useful discussions.

\section{Appendix}

\appendix

\section{The symmetries of the three loop ladder skeleton}

In the calculations of the four graviton amplitude that involve only the ladder diagrams, we see that the lattice factor $F_L$, and $\Delta_3$ which arises from integrating over the non--compact momenta have some symmetries. These correspond to the freedom of relabelling the loop momenta in the integrals (and relabelling the KK momenta in the infinite sums). We find it very useful for our purposes to express the amplitudes in a manifestly symmetric way. This generalizes the analogous arguments for the amplitude at two loops as well as the three loop amplitude arising from the Mercedes diagrams. Parametrizing the ladder skeleton in terms of the Schwinger parameters is expressed in figure 9. Thus though there are 6 links in the skeleton diagram, it is described by only 5 Schwinger parameters as follows from the momentum flow in the diagram\footnote{In fact the number of Schwinger parameters increases by 2 as one increases the number of loops by 1 for the ladder skeleton diagrams.}.  

\begin{figure}[ht]
\begin{center}
\[
\mbox{\begin{picture}(240,110)(0,0)
\includegraphics[scale=.5]{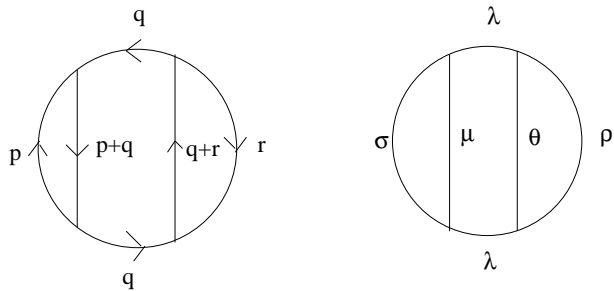}
\end{picture}}
\]
\caption{Parametrizing the three loop ladder skeleton}
\end{center}
\end{figure}

It is easy to see the symmetries from the dual diamond graph, which follows from replacing each face of the ladder skeleton with a vertex of the diamond graph, such that every edge of the diamond graph is parametrized by the link of the ladder skeleton that it cuts as depicted in figure 10. Thus the symmetry group is the set of discrete transformations which interchanges the 4 vertices of either diagram, keeping the links between the vertices intact. This is the Klein four--group $K_4$ which has four elements.  

\begin{figure}[ht]
\begin{center}
\[
\mbox{\begin{picture}(240,110)(0,0)
\includegraphics[scale=.5]{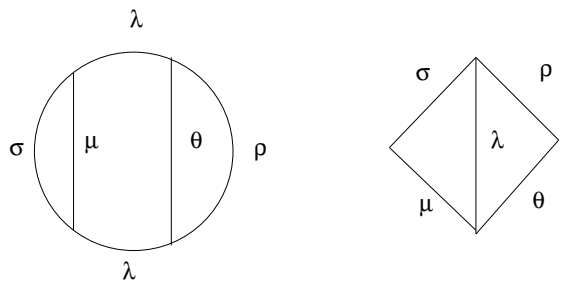}
\end{picture}}
\]
\caption{The three loop ladder skeleton and the dual diamond graph}
\end{center}
\end{figure}

\subsection{Transformations of the Schwinger parameters under $K_4$}

It is easy to see how the 5 Schwinger parameters transform under the action of $K_4$. We now write down the action of the transformations of $K_4$ on the 5 Schwinger parameters. We shall use the notation

\be P_i = \begin{pmatrix}
 \s &\lambda &\rho &\mu &\theta \\
* & * & *& *& *
\end{pmatrix}\ee
where $i=1,\cdots, 4$ to indicate each transformation, where the lower row stands for the parameters $\s_i (\s ~\lambda ~\rho ~\mu ~\theta)$, where $\s_i \in K_4$. Thus the 4 transformations are given by
\bea \label{listtrans}P_1 = \begin{pmatrix}
\s &\lambda &\rho &\mu &\theta \\
\s &\lambda &\rho &\mu &\theta 
\end{pmatrix} \quad
P_2 = \begin{pmatrix}
\s &\lambda &\rho &\mu &\theta \\
\rho &  \lambda& \s& \theta& \mu
\end{pmatrix}  \non \\
P_3 = \begin{pmatrix}
\s &\lambda &\rho &\mu &\theta \\
\mu & \lambda & \theta& \s& \rho
\end{pmatrix}\quad
P_{4}=\begin{pmatrix}
\s &\lambda &\rho &\mu &\theta \\
\theta & \lambda & \mu& \rho& \s
\end{pmatrix} .
\eea 
Now $\Delta_3 (\s,\lambda,\rho,\mu,\theta), F_L (\s,\lambda,\rho,\mu,\theta)$ and $d\Upsilon$ are $K_4$ invariants.

\section{The contribution from $T_1 =0$ in \C{mostimp}}

We consider the point $T_1=0$ in \C{mostimp}. Then \C{constraint} reduces to 
\be 0 \leq \Theta_1 \leq 1, \quad \Theta_2 \geq 0, \quad \Big\vert \Theta -\frac{1}{2}\Big\vert^2 \geq \frac{1}{4} ,\ee
where
\be \Theta = A +i\sqrt{\frac{L^3}{T_2}}.\ee
Hence we see that $\Theta$ lies in the region $\mathcal{R}_\Theta =f_1 \oplus f_2 \oplus g_1 \oplus g_2 \oplus h_1 \oplus h_2$ in figure 11.

\begin{figure}[ht]
\begin{center}
\[
\mbox{\begin{picture}(150,110)(0,0)
\includegraphics[scale=.55]{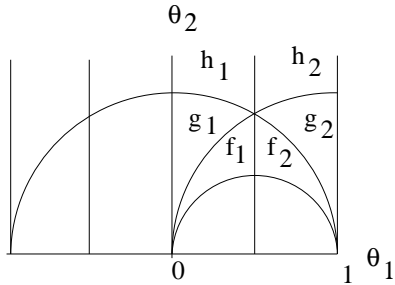}
\end{picture}}
\]
\caption{The $\Theta$ plane}
\end{center}
\end{figure}

The inverse metric in the lattice sum is given by
\bea G^{\alpha\beta} = l_{11}^2 V_3^{-2/3} L^{-2}\begin{pmatrix}
1 & A & 0  \\
A & A^2 + L^3/T_2 & 0 \\
0 & 0 &  L^3 T_2
\end{pmatrix} \non \\ = l_{11}^2 V_3^{-2/3} \lambda^{-1} \begin{pmatrix}
1/\Theta_2 & \Theta_1/\Theta_2 & 0  \\
\Theta_1/\Theta_2 & \vert \Theta \vert^2 /\Theta_2 & 0 \\
0 & 0 &  \lambda^3
\end{pmatrix},\eea
where 
\be \lambda = \sqrt{L T_2}\ee
which has an $SL(2,\mathbb{Z})_\Theta$ invariant subspace. Thus we have that
\bea 
I^{D^8\mathcal{R}^4} = \frac{\pi^{21/2}\Sigma_4}{64 l_{11}^{13}} \int_0^\infty d V_3 V_3^{10/3} \int_0^\infty \frac{d\lambda}{\lambda^2} \int_{\mathcal{R}_\Theta} \frac{d\Theta_1 d\Theta_2}{\Theta_2^2} \hat{F}_L (\lambda,\Theta) \frac{(1-\Theta_1)(\vert \Theta\vert^2 - \Theta_1)}{\Theta_2^2},\non \\ \eea 
where
\be \hat{F}_L (\lambda,\Theta) \equiv \hat{F}_L(L,0,T_2,A).\ee

As in the analysis for the $A=0$ case in \C{impmost3}, we now map $\mathcal{R}_\Theta$ into $\mathcal{F}_\Theta = F \oplus F'$ (see figure 7) keeping track of how the integrand changes when the 6 patches in $\mathcal{R}_\Theta$ map into the 2 patches of $\mathcal{F}_\Theta$. Note that the various lattice factors are the same as they can all be related by $SL(2,\mathbb{Z})_\Theta$ transformations. As the analysis is very similar to the analysis before, we only mention the results.  

{\bf{(i)}} The regions $h_1$ and $g_1$ are mapped by $SL(2,\mathbb{Z})_\Theta$ transformations in the following way:
\bea &&h_1 \rightarrow F, \quad \Theta' =\Theta, \non \\ &&g_1 \rightarrow F' , \quad \Theta' = -1/\Theta. \eea
Their total contribution to the $\Theta$ integral is
\be \label{part1}\int_{\mathcal{F}_\Theta} \frac{d\Theta_1 d\Theta_2}{\Theta_2^2} \hat{F}_L (\lambda,\Theta) \frac{(1-\vert \Theta_1 \vert)(\vert \Theta\vert^2 - \vert \Theta_1 \vert)}{\Theta_2^2}.\ee

{\bf{(ii)}} The regions $g_2$ and $f_2$ are mapped by $SL(2,\mathbb{Z})_\Theta$ transformations in the following way:
\bea &&g_2 \rightarrow F, \quad \Theta' =1/(1-\Theta), \non \\ &&f_2 \rightarrow F' , \quad \Theta' = \Theta/(1-\Theta). \eea
Their total contribution to the $\Theta$ integral is
\be \label{part2}\int_{\mathcal{F}_\Theta} \frac{d\Theta_1 d\Theta_2}{\Theta_2^2} \hat{F}_L (\lambda,\Theta) \frac{\vert \Theta_1 \vert(1-\vert \Theta_1 \vert)}{\Theta_2^2}.\ee

{\bf{(iii)}} The regions $f_1$ and $h_2$ are mapped by $SL(2,\mathbb{Z})_\Theta$ transformations in the following way:
\bea &&f_1 \rightarrow F, \quad \Theta' =(\Theta -1)/\Theta, \non \\ &&h_2 \rightarrow F' , \quad \Theta' = \Theta -1. \eea
Their total contribution to the $\Theta$ integral is
\be \label{part3}\int_{\mathcal{F}_\Theta} \frac{d\Theta_1 d\Theta_2}{\Theta_2^2} \hat{F}_L (\lambda,\Theta) \frac{\vert \Theta_1 \vert(\vert \Theta\vert^2 - \vert \Theta_1 \vert)}{\Theta_2^2}.\ee

On adding the contributions \C{part1}, \C{part2} and \C{part3}, we see that the complete amplitude is given by
\bea \label{t0}
I^{D^8\mathcal{R}^4} = \frac{\pi^{21/2}\Sigma_4}{64 l_{11}^{13}} \int_0^\infty d V_3 V_3^{10/3} \int_0^\infty \frac{d\lambda}{\lambda^2} \int_{\mathcal{F}_\Theta} \frac{d\Theta_1 d\Theta_2}{\Theta_2^2} \hat{F}_L (\lambda,\Theta). \eea 
The integral over $\mathcal{F}_\Theta$ is $SL(2,\mathbb{Z})_\Theta$ invariant. 

We now evaluate the finite part of \C{t0}. We have that
\be I^{D^8\mathcal{R}^4} = b_0 \Lambda^{13} +\ldots + (l_{11}^2 \mathcal{V}_2)^{-13/2} g(\Omega,\bar\Omega),\ee
where $b_0$ is a moduli independent constant, $g(\Omega,\bar\Omega)$ is the $SL(2,\mathbb{Z})$ invariant finite part of $I^{D^8\mathcal{R}^4}$ and the remaining contributions involve various one loop and two loop divergences.

Now in the expression for $\hat{F}_L$ the terms involving $\hat{n}^I$ separate from the terms involving $\hat{l}^I, \hat{m}^I$.  Hence defining
\be V_3^{2/3} \lambda^{-2} = x, \quad V_3^{2/3} \lambda = y,\ee
the $x$ integral which involves $\hat{n}^I$ can be easily done to yield
\bea \label{T0}
I^{D^8\mathcal{R}^4} = \frac{3\pi^6\Sigma_4}{512 (l_{11}^2 \mathcal{V}_2)^{13/2}} E_{5/2} (\Omega,\bar\Omega) \int_0^\infty d y  y ^3  \int_{\mathcal{F}_\Theta} \frac{d\Theta_1 d\Theta_2}{\Theta_2^2}  \sum_{\hat{m}^I,\hat{n}^I} e^{-\pi^2 y \hat{G}_{IJ} (\hat{m} +\hat{n}\Theta)^I(\hat{m}+\hat{n}\bar{\Theta})^J/\Theta_2}. \non \\ \eea 

We now evaluate the finite part that results from the remaining part of the expression in \C{T0}. Since the $\Theta$ integral is $SL(2,\mathbb{Z})$ invariant, our analysis is along the lines of~\cite{Green:1999pu} on using the results in~\cite{Dixon:1990pc}. The essential point is that using the $SL(2,\mathbb{Z})$ invariance, the lattice sum splits into contributions from the zero, degenerate and non--degenerate orbits of $SL(2,\mathbb{Z})$. While the zero and degenerate orbits yield divergent contributions, the non--degenerate orbits yield the finite contribution. This contribution is given by 
\bea 
&& I^{D^8\mathcal{R}^4}_{finite} \non \\ && = \frac{3\pi^6\Sigma_4}{512 (l_{11}^2 \mathcal{V}_2)^{13/2}} E_{5/2} (\Omega,\bar\Omega) \int_0^\infty d y  y ^3  \int_{\mathcal{F}_\Theta} \frac{d\Theta_1 d\Theta_2}{\Theta_2^2}  \sum_{\hat{m}^I,\hat{n}^I} e^{-\pi^2 y \hat{G}_{IJ} (\hat{m} +\hat{n}\Theta)^I(\hat{m}+\hat{n}\bar{\Theta})^J/\Theta_2}\Big\vert_{non-degen} \non \\&& = \frac{3\pi^6\Sigma_4}{256 (l_{11}^2 \mathcal{V}_2)^{13/2}} E_{5/2} (\Omega,\bar\Omega) \int_0^\infty d y  y ^3 \int_{-\infty}^\infty d S_1 \int_0^\infty \frac{dS_2}{S_2^2} \sum_{0 \leq j < k, p \neq 0} e^{2\pi^2 ykp -\pi^2 y\vert kS + j + p\Omega\vert^2/S_2\Omega_2}.\non \\ \eea 

Now
\bea \label{int}&&\int_0^\infty d y  y ^3 \int_{-\infty}^\infty d S_1 \int_0^\infty \frac{dS_2}{S_2^2} \sum_{0 \leq j < k, p \neq 0} e^{2\pi^2 ykp -\pi^2 y\vert kS + j + p\Omega\vert^2/S_2\Omega_2} \non \\ &&= \sqrt{\frac{\Omega_2}{\pi}} \sum_{0 \leq j < k, p \neq 0} \frac{1}{k}\int_0^\infty d y  y ^3 \int_0^\infty \frac{dS_2}{S_2^2} \sqrt{\frac{S_2}{y}} e^{-\pi^2 y(k^2 S_2^2 + p^2 \Omega_2^2)/S_2\Omega_2}  \non \\&&= \sqrt{\frac{\Omega_2}{\pi}} \sum_{k=1}^\infty \int_0^\infty dz\sqrt{z} e^{-\pi^2k^2z/\Omega_2} \sum_{p=1}^\infty  \int_0^\infty dx x e^{-\pi^2 p^2\Omega_2 x} \non \\ &&= \frac{\zeta(3)\zeta(4)}{2\pi^7}.\eea

In performing the integral in \C{int}, we have first integrated over $S_1$, and then we have substituted
\be z = yS_2, \quad x= y/S_2.\ee
Note that the integral is independent of $\Omega$.

Thus we get the finite contribution at $T_1 =0$ given by
\be I^{D^8\mathcal{R}^4} = \frac{3 \zeta(3) \zeta(4)\Sigma_4}{512 \pi (l_{11}^2 \mathcal{V}_2)^{13/2}} E_{5/2} (\Omega,\bar\Omega). \ee

Thus upto an overall numerical factor, this leads to a term in the 9 dimensional effective action
\be l_{11}^7 \int d^9 x \sqrt{-G^{(9)}} \mathcal{V}_2^{-11/2} E_{5/2} (\Omega,\bar\Omega) D^8\mathcal{R}^4.\ee
Again in the IIB theory, this leads to
\be \frac{8}{3} \zeta (4) l_s^7 \int d^9 x \sqrt{-g^{B}} r_B^5 e^{2\phi^B} D^8\mathcal{R}^4,\ee
where we have ignored a genus zero contribution proportional to $r_B^5 \zeta(5) e^{-2\phi^B}.$
This leads to a possible local contribution at genus 2 in string theory.   

\section{Relating M theory on $T^2$ to the type II theories}

We briefly mention the various relations expressing quantities in M theory on $T^2$ in terms of quantities describing the type IIB theory on $S^1$~\cite{Hull:1994ys,Witten:1995ex,Aspinwall:1995fw,Schwarz:1995jq}. The eleven dimensional Planck length $l_{11}$ is related to the string length $l_s$ by the relation
\be l_{11} = e^{\phi^B/3} r_B^{-1/3}l_s,\ee
where $\phi^B$ is the type IIB dilaton and $r_B$ is the radius of the $S^1$ in the string frame. 
The volume $\mathcal{V}_2$ (in units of $4\pi^2 l_{11}^2$) and complex structure $\Omega$ of $T^2$ are related to quantities in the type IIB theory by
\be \mathcal{V}_2 = e^{\phi^B/3} r_B^{-4/3}, \quad \Omega_1 = C, \quad \Omega_2 = e^{-\phi^B},\ee
where $C$ is the 0 form potential. To obtain expressions in the type IIA theory, we use the relations
\be \label{relate}r_A r_B = 1, \quad e^{-\phi^A} = r_B e^{-\phi^B} \ee
where $\phi^A$ is the type IIA dilaton and $r_A$ is the radius of the $S^1$ in the string frame. Also $C$ is the 1 form potential reduced on the $S^1$. 

\section{The coefficients $a_1, b_1, c_1, d_1$ and $e_1$}

We now consider the equations \C{A1} and \C{c1}. As discussed in the main text, consistency of the supergravity calculations with perturbative string theory sets $a_0 = b_0 = c_0 =d_0 =e_0 =0$.
Thus these Poisson equations are of the form
\be \label{mult} [\Delta_\Omega - s(s-1) ]f (\Omega,\bar\Omega)= g (\Omega,\bar\Omega),\ee
where
\be f (\Omega,\bar\Omega) = A \Omega_2^{1-s}+\ldots,\ee
where $A$ is the coefficient we want to determine. Now $A$ is not determined by the perturbative structure of $g$, and hence we must use the non--perturbative information in $g$ to determine $A$. To do so, we multiply \C{mult} by $E_s (\Omega,\bar\Omega)$ and integrate over the fundamental domain of $\Omega$ with the $SL(2,\mathbb{Z})$ invariant measure. However this is divergent and we integrate over the truncated fundamental domain $\mathcal{F}_L$, where $\Omega_2 \leq L$, and take $L\rightarrow \infty$. On the left hand side we are left with the boundary contribution
\be \Big( E_s \frac{\p f}{\p \Omega_2} - f\frac{\p E_s}{\p \Omega_2} \Big) \Big\vert_{\Omega_2 = L\rightarrow \infty}.\ee 
 On the right hand side, we express $E_s$ as a Poincare series, and use the Rankin--Selberg formula to unfold the integral to get 
\be \label{RS}\int_{\mathcal{F}_L} \frac{d^2\Omega}{\Omega_2^2} E_s g = 2\zeta(2s) \int_0^L d\Omega_2 \Omega^{s-2} \int_{-1/2}^{1/2} d\Omega_1 g,\ee
as $L\rightarrow \infty$. 
The finite contributions from both sides give $A$, while the divergent ones trivially match~\cite{Green:2005ba,Green:2008bf}. In obtaining $A$, use is made of Ramanujan's identity
\be \label{rama}
\sum_{k=1}^\infty \frac{\mu(k,s)\mu(k,s')}{k^r} = \frac{\zeta(r)\zeta(r+2s-1)\zeta(r+2s'-1)\zeta(r+2s+2s'-2)}{\zeta(2r+2s+2s'-2)}.\ee
Using this, $c_1$ and $d_1$ are easily determined to be 
\bea c_1 = \frac{8}{6435 \zeta(4)} \sum_{k=1}^\infty \frac{\mu(k,3/2)\mu(k,5/2)}{k^4} =\frac{2\zeta(10)}{1575}, \non \\ 
d_1 = \frac{19}{924\zeta(2)} \sum_{k=1}^\infty \frac{\mu(k,3/2)\mu(k,5/2)}{k^2} =\frac{19\zeta(8)}{840}.\eea

For the other coefficients, the analysis is quite complicated. This is because the relevant integral in \C{RS} over $\Omega_2$ after substituting the expression for $g$ diverges as $\Omega_2 \rightarrow 0$. This is a complication that arises on implementing the Rankin--Selberg formula over the truncated fundamental domain $\mathcal{F}_L$. To schematically see the origin of the complication, note that the original integral was over the regions $F \oplus F'$ in figure 7. Now perform an $SL(2,\mathbb{Z})$ transformation $\Omega \rightarrow -1/\Omega$ such that $\Omega_2 = i\infty$ is mapped to $\Omega_2 =0$. Hence $F\oplus F'$ is mapped to the region $\vert\Omega\vert^2 \leq 1, \vert\Omega \pm 1 \vert^2 \geq 1$. Now considering the total contribution from the two regions, unfolding essentially yields \C{RS}. However, note that the divergences at $\Omega_2 =L$ get mapped to those at $\Omega_2 = L^{-1}$. Hence these contributions also have to included. We simply implement this cutoff and calculate the finite part of the integral, to yield a contribution to $A$. We expect this to yield at least a part of the total contribution to $A$ for generic cases\footnote{In fact implementing this cutoff one can check that all the divergences as $L\rightarrow \infty$ do not cancel, hence showing this does not yield the complete answer.}. This complication arises whenever $s \leq s_0$, where $s_0$ is the maximum value where the solution has a logarithm in $\Omega_2$ for given $g$. It would be interesting to obtain the exact expression in such cases.

Hence proceeding as stated above, to partially determine $a_1$ we have that
\be \label{a1}14 a_1 + \frac{128\cdot 80 \zeta(4)^2}{63\cdot 81}= -\frac{320(4\pi^{5/2})^2}{81 \Gamma(5/2)^2} \sum_{k=1}^\infty k^4 \mu^2 (k,5/2) \int_{1/L}^\infty d\Omega_2 \Omega_2^3 K_2^2 (2\pi k\Omega_2).\ee
The integral, which can be expressed in terms of a Meijer G--function, gives us a finite part equal to
\be \label{a11} \frac{1}{(2\pi k)^4} \Big[ \frac{1}{3} -4\gamma + {\rm ln}(16/(2\pi)^4) - 4 {\rm ln}k \Big] ,\ee  
where $\gamma$ is the Euler--Mascheroni constant, as $L\rightarrow \infty$. The contribution from the $k$ independent terms in \C{a11} can be easily evaluated using \C{rama}, while that from the ${\rm ln}k$ term is obtained from \C{rama} on differentiating with respect to $r$ and setting $r=0$, leading to
\be \sum_{k=1}^\infty {\rm ln}k \mu^2 (k,5/2) = \zeta(4)^2 \Big[ \frac{1}{2}{\rm ln}2\pi +\frac{\zeta'(4)}{\zeta(4)} - \frac{\zeta'(8)}{2\zeta(8)}\Big] .\ee
Thus we get that
\be 14 a_1 = -\frac{12800}{15309} \zeta(4)^2 +\frac{2560}{729} \zeta(4)^2 \Big[ {\rm ln} (16e^{-4\gamma}) +\frac{8\zeta' (4)}{\zeta(4)} -\frac{4\zeta' (8)}{\zeta(8)}\Big].\ee
We have used the results \be \zeta(0) =-1/2, \quad \zeta'(0) = -\frac{1}{2}{\rm ln}2\pi  \ee
in obtaining the answer.

Similarly to partially determine $b_1$ we have that
\be \label{b1}b_1 = \frac{3(4\pi^{5/2})^2}{10 \Gamma(5/2)^2} \sum_{k=1}^\infty k^4 \mu^2 (k,5/2) \int_{1/L}^\infty d\Omega_2 \Omega_2 K_2^2 (2\pi k\Omega_2).\ee
The integral can be expressed in terms of the modified Bessel functions $K_n$ and gives a finite part equal to
\be \frac{1}{(2\pi k)^2} \Big[ -\frac{1}{2} +2\gamma +2{\rm ln}\pi +2{\rm ln} k \Big].\ee 
Proceeding as above, the contribution from the $k$ independent terms vanishes using $\zeta(-2)=0$, while that from the ${\rm ln}k$ term is evaluated using
\be \sum_{k=1}^\infty k^2 {\rm ln}k \mu^2(k,5/2) = \frac{\zeta(2)\zeta(3)\zeta(6)}{24\zeta(4)}\ee 
on using
\be \zeta' (-2) = -\frac{\zeta(3)}{4\pi^2},\ee
which leads to
\be b_1 = \frac{8}{3} \zeta(3) \zeta(6).\ee

Finally to partially determine $e_1$ we have that
\be \label{e1}10 e_1 -\frac{16\zeta(2)\zeta(4)}{35}= \frac{48\pi^4}{7 \Gamma(3/2) \Gamma(5/2)} \sum_{k=1}^\infty k^3 \mu(k,3/2)\mu (k,5/2) \int_{1/L}^\infty d\Omega_2 \Omega_2^2 K_1 (2\pi k\Omega_2) K_2 (2\pi k\Omega_2).\ee
Once again, the integral can be expressed in terms of a Meijer G--function with a finite part equal to
\be \frac{1}{(2\pi k)^3} \Big[ -\frac{1}{2} -2\gamma -2{\rm ln}\pi -2{\rm ln}k \Big].\ee
Proceeding as above and using
\be \sum_{k=1}^\infty {\rm ln}k \mu(k,3/2)\mu(k,5/2) = \frac{1}{2} {\rm ln}2\pi \zeta(2) \zeta(4)+\frac{1}{2} \zeta'(2) \zeta(4) +\frac{1}{2} \zeta(2)\zeta'(4) -\frac{1}{2} \zeta(2)\zeta(4)\frac{\zeta'(6)}{\zeta(6)},\ee
we get that
\be 10 e_1 = \frac{36}{35} \zeta(2) \zeta(4) +\frac{16}{7} \zeta(2)\zeta(4) \Big[ {\rm ln}(e^\gamma/2) -\frac{\zeta'(2)}{\zeta(2)} -\frac{\zeta'(4)}{\zeta(4)} +\frac{\zeta'(6)}{\zeta(6)}\Big].\ee
 
Note that similar manipulations occur in~\cite{Green:2008bf}. However, there are no contributions which diverge at the lower end of the $\Omega_2$ integral like ours. This is because $v^1 = f^2_{(3,0)}=f^2_{(0,2)}=0$ (see equations (A.29), (A.47) and (A.48)). 

\section{Generalities and our analysis of the two loop amplitude}

In the main text, we have evaluated the three and four loop amplitudes for certain values of the moduli. We now give a justification for why these special choices of the moduli should play an important role in determining the structure of these amplitudes.   

In the absence of concrete answers at three and four loops, we consider in detail the analysis at two loops, where several amplitudes are known~\cite{Green:1999pu,Green:2005ba,Green:2008bf} and we can use the results to compare with what we get based on our analysis in the main text. This analysis is entirely analogous to what we have done at three and four loops. 

At two loops the auxiliary geometry is known to be $T^2$, and one integrates over the volume $V_2$ and complex structure $T$ of $T^2$ over a certain range of the parameters. Thus the metric of the auxiliary $T^2$ is
\be 
G_{\alpha\beta} = \frac{V_2}{T_2} \left( \begin{array}{cc} \vert T \vert^2 & -T_1 \\ -T_1 & 1 \end{array} \right)\ee
and one can get the exact answer for the various amplitudes. What if we did not know about this underlying geometry, and looked at the amplitudes when $T_1 =0$? This is a case when the metric diagonalizes and the calculation is simpler. This is analogous to what we have done at higher loops. Physically for such a choice, the lattice factor 
\be \sum_{\hat{k}^{\alpha I}} e^{-\pi^2 G_{\alpha\beta} G_{IJ} \hat{k}^{\alpha I} \hat{k}^{\beta J}}\ee 
factorizes into U--duality invariant products of lattice factors associated with the diagonal blocks of the metric $G_{\alpha\beta}$. Along with the various other terms in the integrand, we obtain U--duality invariant expressions on integrating over the remaining part of moduli space. 

Now note that in performing this analysis, we have split the metric $G_{\alpha\beta}$ into diagonal blocks, each of which leads to expressions for the amplitude involving lower loops in supergravity involving moduli from that block only. Hence these produce U--duality invariant contributions to the amplitude from lower loops in supergravity. These lead to precisely the ``source terms'' in the Poisson equations that the moduli dependent coefficients of these interactions satisfy~\cite{Basu:2008cf}. Hence we generically expect to reproduce the structure of various source terms from our analysis. The main thrust of the analysis is to see how much of these source terms we can determine. 

So how do the two different analysis tally together? It is not difficult to see that they yield the same structure so far as the perturbative and some parts of the non-perturbative parts of the amplitude are concerned, including the transcendentality. For example, suppose that our analysis for special values of the moduli yield a perturbative contribution to the amplitude of the form
\be a_0 \Omega_2^{p_0} + a_1 \Omega_2^{p_1} +\ldots. \ee 
when expanded at weak coupling. Hence we expect that the perturbative part of the exact expression $f$ for the amplitude will involve a Poisson equation of the form\footnote{In general we expect $f$ to split into a sum of modular forms each of which satisfies a Poisson equation. This detail is irrelevant for our analysis.}
\be \Omega_2^2 \frac{\p^2 f}{\p\Omega_2^2} = \ldots + (a_0 \Omega_2^{p_0} + a_1 \Omega_2^{p_1} +\ldots).\ee 
Clearly the $\Omega_2$ dependence and the transcendentality is captured by our analysis. We now consider the details of the two loop amplitudes to illustrate our point. We shall drop all irrelevant numerical factors and keep only those needed.  
   
The $D^4\mathcal{R}^4$ interaction at two loops is given by~\cite{Green:1999pu}
\be \frac{\pi^{11}}{2l_{11}^8} \sum_{m_I,n_I} \int_0^\infty dV_2 V_2^3 \int_{\mathcal{F}_T} \frac{dT_1 dT_2}{T_2^2} e^{-\pi^2 G_{IJ}(m+nT)_I(m+n\bar{T})_J V_2/T_2},\ee 
and the finite part is equal to
\be \label{case11}\frac{\pi^4 \zeta(3)\zeta(4)}{2l_{11}^8 \mathcal{V}_2^4}.\ee

To illustrate what our analysis yields, we start with the expression 
\be \frac{\pi^{11}}{6l_{11}^8} \sum_{m_I,n_I} \int_0^\infty dV_2 V_2^3 \int_{D_T} \frac{dT_1 dT_2}{T_2^2} e^{-\pi^2 G_{IJ}(m+nT)_I(m+n\bar{T})_J V_2/T_2}\ee 
one obtains from the two loop diagrams
where $D_T$ is the region
\be 0 \leq T_1 \leq 1, \quad \Big\vert T - \frac{1}{2} \Big\vert^2 \geq \frac{1}{4}.\ee
Setting $T_1=0$, the finite part is given by
\be \frac{\pi^4 E_{3/2} E_{5/2}}{32l_{11}^8 \mathcal{V}_2^4} \ee 
leading to perturbative contributions
\be \label{case12}\frac{\pi^4}{2 l_{11}^8 \mathcal{V}_2^4} \Big( \frac{\zeta(3)\zeta(4)}{3} + \frac{2\zeta(2)\zeta(4)}{3} \Omega_2^{-2} + \frac{\zeta(2)\zeta(5)}{2} \Omega_2^2 + \frac{\zeta(3)\zeta(5)}{4} \Omega_2^4\Big).\ee
The structure of the term independent of $\Omega_2$ matches in \C{case11} and \C{case12}, while the rest does not (one can show that the rest do not contribute to the perturbative part of the amplitude). This is not unexpected as there are no source terms for the $D^4\mathcal{R}^4$ interaction~\cite{Basu:2008cf} and so block diagonalizing the metric to get the contributions is not particularly useful. 

Now let us consider the $D^6\mathcal{R}^4$ interaction, where the total contribution is given by~\cite{Green:2005ba}
\be \frac{\pi^{11}}{12 l_{11}^6}\sum_{m_I,n_I} \int_0^\infty dV_2 V_2^2 \int_{\mathcal{F}_T} \frac{dT_1 dT_2}{T_2^2} e^{-\pi^2 G_{IJ}(m+nT)_I(m+n\bar{T})_J V_2/T_2} A(T,\bar{T}),\ee
where
\be A(T,\bar{T}) = \frac{\vert T \vert^2 -\vert T_1 \vert +1}{T_2} +\frac{5}{T_2^3}(T_1^2 - \vert T_1 \vert)(\vert T \vert^2 - \vert T_1 \vert).\ee
The finite part of the amplitude 
\be \frac{\pi^6 \mathcal{E}}{96 l_{11}^6 \mathcal{V}_2^3}\ee
satisfies the Poisson equation
\be \label{case21}4\Omega_2^2 \frac{\p^2\mathcal{E}}{\p\Omega \p\bar\Omega} = 12 \mathcal{E} -6 E_{3/2}^2.\ee
Hence, the finite part of the amplitude receives perturbative contributions
\be \label{case211}\frac{\pi^6}{24 l_{11}^6 \mathcal{V}_2^3} \Big( \zeta(3)^2 \Omega_2^3 + 2\zeta(2)\zeta(3)\Omega_2 + 6\zeta(4)\Omega_2^{-1} + \frac{2}{9}\zeta(6)\Omega_2^{-3}\Big).\ee
Notably $A(T,\bar{T})$ satisfies the Poisson equation
\be \label{case210}4 T_2^2 \frac{\p^2 A}{\p T \p \bar{T}} = 12 A - 12 T_2 \delta(T_1).\ee
To perform our analysis, we start with
\be \frac{\pi^{11}}{36 l_{11}^6}\sum_{m_I,n_I} \int_0^\infty dV_2 V_2^2 \int_{D_T} \frac{dT_1 dT_2}{T_2^2} e^{-\pi^2 G_{IJ}(m+nT)_I(m+n\bar{T})_J V_2/T_2} A(T,\bar{T}).\ee
Again setting $T_1 =0$ and hence setting $A = T_2 + T_2^{-1}$, we see that the finite part is given by
\be \label{case22}\frac{\pi^6}{288 l_{11}^6 \mathcal{V}_2^3} \Big(E_{3/2}^2 + 3 E_{1/2} E_{5/2}\Big).\ee
The structure of the $E_{3/2}^2$ term agrees in \C{case21} and \C{case22}, while the other term does not\footnote{In fact the source term $E_{3/2}^2$ in \C{case21} can be argued based on the structure of supersymmetry~\cite{Basu:2008cf} . There is no other source term for this interaction.}. Now the $E_{3/2}^2$ term in \C{case22} yields the perturbative contribution
\be  \label{pcont}\frac{\pi^6}{72 l_{11}^6 \mathcal{V}_2^3} \Big( \zeta(3)^2 \Omega_2^3 + 4\zeta(2)\zeta(3)\Omega_2 + 10\zeta(4)\Omega_2^{-1} \Big)\ee
to the amplitude. Note that this has a very similar structure to \C{case211}, including the $\Omega_2$ dependence and transcendentality (the $\Omega_2^{-3}$ term in \C{case211} does not come from the source term). 
 
In fact we can immediately see the reason for the difference in the structures of \C{case21} and \C{case22}. In the exact expression, the source term contribution to the integrand (at $T_1 =0$) is proportional to $T_2$ on using \C{case210}, while we have a contribution proportional to $T_2 + T_2^{-1}$, leading to the extra term of the form $E_{1/2}E_{5/2}$.

What about the non--perturbative contributions?  The total contribution to the $D^6\mathcal{R}^4$ interaction can be divided into two parts: (i) contributions carrying vanishing D--(anti)instanton charge and (ii) contributions carrying non--vanishing charge. 

First let us consider what our analysis yields. Contributions (i) include the perturbative contributions given in \C{pcont} and the non--perturbative contributions
\be \frac{2\pi^8 \Omega_2}{9 l_{11}^6 \mathcal{V}_2^3} \sum_{n\neq 0} n^2 \mu^2 (\vert n\vert, 3/2) K_1^2 (2\pi\vert n\vert\Omega_2)\ee
obtained from keeping only $E_{3/2}^2$ in \C{case22} on using \C{Eisenstein}. This includes all contributions from instanton--anti--instanton bound states carrying total vanishing instanton charge. The leading contributions in the weak coupling expansion are given by
\be \label{Match}\frac{\pi^8}{9l_{11}^6 \mathcal{V}_2^3} \sum_{n=1}^\infty n \mu^2 (n,3/2) e^{-4\pi n\Omega_2} \Big(1 + \frac{3}{8\pi n\Omega_2}- \frac{3}{128 \pi^2 n^2 \Omega_2^2}+O((n \pi\Omega_2)^{-3})\Big).\ee

On the other hand, the exact analysis yields the perturbative contributions given by \C{case211}, and the leading non--perturbative contribution at weak coupling is given by~\cite{Green:2014yxa}
\be \label{Match2}-\frac{3\pi^8}{4l_{11}^6 \mathcal{V}_2^3} \sum_{n=1}^\infty n \mu^2 (n,3/2)e^{-4\pi n \Omega_2} \Big( \frac{1}{n^2 \pi^2 \Omega_2^2} +O((n\pi\Omega_2)^{-3})\Big) .\ee
From the general structure of the expressions, it follows that apart from the first two terms in \C{Match}, the structure of the terms in \C{Match2} matches those in \C{Match} apart from numerical factors (this is also true for an infinite class of terms if one keeps all the subleading terms). Hence in the zero charge instanton sector, our analysis does capture several features of the exact answer.

Next we consider what our analysis yields to contributions (ii). The total contribution is given by
\be \sum_{n\neq 0}\tilde{\mathcal{E}}_n (\Omega_2) e^{2\pi i n\Omega_1}\ee 
where the sum involves terms carrying instanton charge $n$. We have that
\bea \label{Eqninst}&&\tilde{\mathcal{E}}_n (\Omega_2) = \frac{\pi^7}{9 l_{11}^6 \mathcal{V}_2^3} \Big[ \Big(\zeta(3) \Omega_2^2 + 2\zeta(2) \Big) \vert n \vert \mu (\vert n \vert, 3/2) K_1 (2\pi \vert n \vert \Omega_2) \non \\ &&+ 2\pi\Omega_2 \sum_{m_i \neq 0, m_1 + m_2 =n} \vert m_1 m_2 \vert \mu (\vert m_1 \vert, 3/2)  \mu (\vert m_2 \vert, 3/2) K_1 (2\pi \vert m_1 \vert \Omega_2) K_1 (2\pi \vert m_2 \vert \Omega_2)\Big].\non \\ \eea

Let us consider the leading perturbative contributions to $\tilde{\mathcal{E}}_n$ in the $n$ instanton charge sector coming from the first term in \C{Eqninst}. They are given by
\bea \label{Match3}\frac{\pi^7 \sqrt{\vert n\vert} \mu (\vert n \vert, 3/2)e^{-2\pi \vert n \vert \Omega_2}}{18 l_{11}^6 \mathcal{V}_2^3 \sqrt{\Omega_2}}\Big[ \zeta(3) \Omega_2^2 + \frac{3\zeta(3)}{16\vert n\vert \pi}\Omega_2 + 2\zeta(2) - \frac{15\zeta(3)}{512 \pi^2 n^2} \non \\ +\Big( \frac{3\zeta(2)}{8\pi\vert n\vert} +\frac{105\zeta(3)}{8192\pi^3 \vert n\vert^3}\Big)\frac{1}{\Omega_2}+O(\Omega_2^{-2})\Big].\eea

One can obtain the similar contributions coming from the exact analysis~\cite{Green:2014yxa} which arise from source terms involving only one Bessel function. The weak coupling expansion  yields\footnote{There is another term which has integer powers of $\Omega_2$ rather than half integer powers as in \C{Match3}. This structure is not reproduced by our analysis.}
\bea \label{Match4}\frac{\pi^7 \sqrt{\vert n\vert} \mu (\vert n \vert, 3/2)e^{-2\pi \vert n \vert \Omega_2}}{3 l_{11}^6 \mathcal{V}_2^3 \sqrt{\Omega_2}}\Big[ \frac{9\zeta(3)}{\vert n\vert \pi}\Omega_2 + \frac{855\zeta(3)}{16 \pi^2 n^2} -\Big( \frac{6\zeta(2)}{\pi\vert n\vert} -\frac{51345\zeta(3)}{512\pi^3 \vert n\vert^3}\Big)\frac{1}{\Omega_2}+O(\Omega_2^{-2})\Big].\non \\ \eea
Apart from the first and third terms in \C{Match3}, the structure of the remaining terms precisely matches between \C{Match3} and \C{Match4}.

Finally consider the contribution arising from the term in the second line of \C{Eqninst} to the $n$ instanton charge sector. This is the sector which involves the most intricate contributions, and we restrict ourselves to the case where $m_1 =m_2 = n/2$ to compare with the exact answer, which is a very simple setting. Our analysis gives a contribution
\be \frac{\pi^8}{36l_{11}^6 \mathcal{V}_2^3}  \vert n\vert \mu^2 (\vert n\vert/2,3/2) e^{-2\pi \vert n\vert\Omega_2} \Big(1 +O(\Omega_2^{-1})\Big).\ee  
This can be compared with non--perturbative contributions involving squares of Bessel functions from the source terms in the exact analysis. Among other contributions, the exact answer indeed has a contribution of the form above given by~\cite{Green:2014yxa}
\be  \frac{\pi^8}{l_{11}^6 \mathcal{V}_2^3}  \vert n\vert \mu^2 (\vert n\vert/2,3/2) e^{-2\pi \vert n\vert\Omega_2} \Big(1 +O(\Omega_2^{-1})\Big).\ee
Thus we see that certain structures of the exact non--perturbative contributions are reproduced by our analysis at all orders in the perturbative expansion around the D--(anti)instanton background.

Let us now extend our analysis beyond the $D^6\mathcal{R}^4$ interaction. We now show that this pattern continues at higher orders in the derivative expansion and we give only the answers (see~\cite{Green:2008bf} for the various expressions given below). For the $D^8\mathcal{R}^4$ term the source terms are of the form
\be \frac{E_{3/2} E_{1/2}}{l_{11}^4 \mathcal{V}_2^2}\ee
and come from a term in the integrand of the form $\delta(T_1)  T_2(T_2 + T_2^{-1})$.
In our analysis, setting $T_1 =0$, we have 
\be  \pi^3 l_{11}^{-4}\sum_{m_I,n_I} \int_0^\infty dV_2 V_2 \int_0^\infty \frac{dT_2}{T_2^2} e^{-\pi^2 G_{IJ}(m_I m_J T_2 + n_I n_J T_2^{-1}) V_2} \Big( 1 + \frac{4}{5} (T_2^2 + T_2^{-2})\Big),\ee
leading to an extra contribution from $T_2^{-2}$ in the integrand. This gives the contribution
\be \frac{3}{20 l_{11}^4 \mathcal{V}_2^2} \Big( 3 E_{3/2} E_{1/2} +\pi^{-2} E_{3/2} E_{5/2}\Big) \ee
where, and later, we have used 
\be \Gamma (s) E_s = \pi^{2s-1} \Gamma (1-s) E_{1-s}.\ee
As before the $E_{3/2}E_{1/2}$ part of the answer matches the structure one obtains from the exact expression.

For the $D^{10} \mathcal{R}^4$ term, the source terms are of the form
\be \frac{1}{l_{11}^2 \mathcal{V}_2} \Big( E_{3/2}^2 +\zeta(2) E_{1/2}^2 \Big)  \ee
which come from a term in the integrand of the form $\delta(T_1)  T_2(1 + T_2^2 + T_2^{-2})$.
On the other hand setting $T_1 =0$, our analysis yields
\be  \pi^3 l_{11}^{-2}\sum_{m_I,n_I} \int_0^\infty dV_2  \int_0^\infty \frac{dT_2}{T_2^2} e^{-\pi^2 G_{IJ}(m_I m_J T_2 + n_I n_J T_2^{-1}) V_2} \Big( \frac{45}{2} (T_2^3 + T_2^{-3}) + 35(T_2 + T_2^{-1})\Big),\ee
leading to an extra contribution from $T_2^{-3}$ in the integrand. This gives the contribution
\be \frac{3}{2l_{11}^2 \mathcal{V}_2} \Big( \frac{35}{8} E_{3/2}^2 + 70\zeta(2) E_{1/2}^2 + \frac{135}{32 \pi^2} E_{5/2}^2 \Big)\ee
whose first two terms match the structure of the exact answer.

Finally for the $D^{12} \mathcal{R}^4$ interaction, the source terms are of the form
\be E_{3/2} E_{5/2} +\zeta(2) E_{1/2} E_{3/2}\ee
which come from a term in the integrand of the form $\delta(T_1)  T_2(T_2 + T_2^{-1})(1 + T_2^2 + T_2^{-2})$. Our analysis at $T_1 =0$ yields
\be \pi^3 \sum_{m_I,n_I} \int_0^\infty dV_2 V_2^{-1} \int_0^\infty \frac{dT_2}{T_2^2} e^{-\pi^2 G_{IJ}(m_I m_J T_2 + n_I n_J T_2^{-1}) V_2} \Big( 56 + 45 (T_2^2 + T_2^{-2}) + 24(T_2^4 + T_2^{-4})\Big),\ee
leading to an extra contribution from $T_2^{-4}$ in the integrand. This gives the contribution
\be \frac{3}{2} \Big( \frac{69}{8} E_{3/2} E_{5/2} + 101 \zeta (2) E_{1/2} E_{3/2} + \frac{45}{4\pi^2} E_{5/2} E_{7/2}\Big)\ee
whose first two terms match the structure of the exact answer. 

Hence our analysis reproduces the structure of the source terms in the Poisson equations the various couplings satisfy, though we get extra contributions. We expect this to generalize to higher loops leading to non--trivial source terms for the non--BPS interactions we have discussed.

%\vspace{.5cm}

%\bibliographystyle{utphys}
%\bibliography{myrefs}
%\providecommand{\href}[2]{#2}\begingroup\raggedright\begin{thebibliography}{10}

\providecommand{\href}[2]{#2}\begingroup\raggedright\endgroup

\end{document}